\documentclass[useAMS,usenatbib]{mn2e}
\usepackage{aecompl}
\usepackage{graphicx}
\usepackage{amssymb}
\usepackage{amsmath}
\usepackage{multicol}
\usepackage{longtable}
\usepackage{longtable}
\usepackage{float}
\usepackage{enumitem}

\usepackage{etoolbox}
\makeatletter
\patchcmd\@combinedblfloats{\box\@outputbox}{\unvbox\@outputbox}{}{%
   \errmessage{\noexpand\@combinedblfloats could not be patched}%
}%
 \makeatother

\usepackage{xcolor,hyperref}

\newcommand{\krome}{\textsc{Krome}}

\title[Non-ideal MHD and microphysics]{The challenges of modelling microphysics: ambipolar diffusion, chemistry, and cosmic rays in MHD shocks}

\author[T. Grassi et al.]{\parbox{\textwidth}{T.~Grassi$^{1,2,3,}$\thanks{Corresponding author: tgrassi@usm.lmu.de}, M.~Padovani$^{3,4}$, J.~P.~Ramsey$^{2,5}$, D.~Galli$^{4}$, N.~Vaytet$^{2,6}$, B.~Ercolano$^{1,3}$, T.~Haugb\o lle$^{2}$}\vspace{0.4cm}\\
\parbox{\textwidth}{
$^{1}$Universit\"ats-Sternwarte M\"unchen, Scheinerstr. 1, D-81679 M\"unchen, Germany\\
$^{2}$Centre for Star and Planet Formation, the Niels Bohr Institute and the Natural History Museum of Denmark, University of Copenhagen, \O stervoldgade 5-7, DK-1350 Copenhagen K, Denmark\\
$^{3}$Excellence Cluster Origin and Structure of the Universe, Boltzmannstr.2, D-85748 Garching bei M\"unchen, Germany\\
$^{4}$INAF-Osservatorio Astrofisico di Arcetri, Largo E. Fermi 5, I-50125 Firenze, Italy\\
$^{5}$Department of Astronomy, University of Virginia, Charlottesville, VA-22904, USA\\
$^{6}$Data Management and Software Centre, the European Spallation Source ERIC, Ole Maal\o es Vej 3, DK-2200, Copenhagen, Denmark\\
}}

\begin{document}

\newcommand{\enzo}{\textsc{Enzo}}
\newcommand{\flash}{\textsc{Flash}}
\newcommand{\ramses}{\textsc{Ramses}}
\newcommand{\fortran}{\textsc{Fortran}}
\newcommand{\python}{\textsc{Python}}
\newcommand{\dlsodes}{\textsc{DLSODES}}
\newcommand{\codename}{\textsc{lemongrab}}
\newcommand{\linktocode}{\url{https://bitbucket.org/tgrassi/lemongrab/}}
\newcommand{\ith}{$i$th }
\newcommand{\jth}{$j$th }
\newcommand{\nth}{$n$th }
\newcommand{\kth}{$k$th }

\newcommand{\dd}{\mathrm d}
\newcommand{\mA}{\mathrm A}
\newcommand{\mB}{\mathrm B}
\newcommand{\mC}{\mathrm C}
\newcommand{\mD}{\mathrm D}
\newcommand{\mE}{\mathrm E}
\newcommand{\mH}{\mathrm H}
\newcommand{\mHe}{\mathrm{He}}
\newcommand{\me}{\mathrm e}
\newcommand{\mSi}{\mathrm Si}
\newcommand{\mO}{\mathrm O}
\newcommand{\mX}{\mathrm X}
\newcommand{\cmc}{\mathrm{cm}^{-3}}
\newcommand{\real}{\mathbb R}
\newcommand{\superscript}[1]{\ensuremath{^{\scriptscriptstyle\textrm{#1}\,}}}
\newcommand{\trader}{\superscript{\textregistered}}

\newcommand{\eqn}[1]{Eq.~(\ref{#1})}
\newcommand{\theireqn}[1]{Eq.~(#1)}
\newcommand{\eqnrange}[2]{Eqs.~(\ref{#1}) to (\ref{#2})}
\newcommand{\sect}[1]{Sect.~\ref{#1}}
\newcommand{\fig}[1]{Fig.~\ref{#1}}
\newcommand{\tab}[1]{Tab.~\ref{#1}}
\newcommand{\appx}[1]{Appendix~\ref{#1}}
\newcommand{\file}[1]{\texttt{#1}}

\newcommand{\arl}[1]{\url{#1}}

\renewcommand\arraystretch{1.7}

\newcommand{\tgcomment}[1]{{\bf #1}}
\newcommand{\jprcomment}[1]{[\textcolor{magenta}{#1 /JPR}]}

\newcommand\osu{osu\_01\_2007}
\def\tm{\leavevmode\hbox{$\rm {}^{TM}$}}

\date{Accepted *****. Received *****; in original form ******}

\pagerange{\pageref{firstpage}--\pageref{lastpage}} \pubyear{2099}

\maketitle

\label{firstpage}

\begin{abstract}
        From molecular clouds to protoplanetary disks, non-ideal magnetic effects are important in many astrophysical environments. Indeed, in star and disk formation processes, it has become clear that these effects are critical to the evolution of the system. The efficacy of non-ideal effects are, however, determined by the complex interplay between magnetic fields, ionising radiation, cosmic rays, microphysics, and chemistry. In order to understand these key microphysical parameters, we present a one-dimensional non-ideal magnetohydrodynamics code and apply it to a model of a time-dependent, oblique, magnetic shock wave. By varying the microphysical ingredients of the model, we find that cosmic rays and dust play a major role, and that, despite the uncertainties, the inclusion of microphysics is essential to obtain a realistic outcome in magnetic astrophysical simulations.
\end{abstract}


\section{Introduction}
Magnetic fields play a key role in determining the structure and evolution of many astrophysical environments. For example, in star forming regions, magnetic fields stabilise against gravity, influence the shape of molecular filaments, and play an important role in large-scale turbulence \citep[see e.g.][]{Padoan2002,McKee2007,Hennebelle2008,Federrath2010}. In protoplanetary disks, magnetic fields can strongly influence the evolution of gas and dust \citep[e.g.][]{Balbus1998,Armitage2011,Flock2016,Xu2016}, remove angular momentum via outflows \citep[e.g.][]{PudritzNorman1983}, affect the dynamical behaviour of planetesimals \citep[e.g.][]{Gressel2011}, and even contribute to heating of (exo)planet atmospheres or to reduce atmospheric loss \citep[e.g.][]{Batygin2013,Cohen2014,Rogers2014,Dong2018}.

How magnetic fields couple to gas and dust in these different contexts depends on the level of ionisation at each location and time. In weakly ionised conditions, such as can be found in molecular clouds and protoplanetary disks, one cannot generally assume that ideal magnetohydrodynamics (MHD) applies. If the magnetic diffusion timescale is comparable to the dynamical timescale, the coupling between magnetic fields and dynamics is regulated by microphysical processes that determine how the ionised matter is ``felt'' by the magnetic fields \citep[e.g.][]{Mestel1956,WardleNg1999,Smith2003,Duffin2008,Tomida2015}. This can significantly affect the structure (e.g.\ outflow launching, disk formation) and dynamics (e.g.\ magnetic braking) of proto-stellar systems (see, e.g., \citealt{Vaytet2018} and references therein).

Unfortunately, there remain significant uncertainties in certain aspects of astrophysically-relevant microphysics and chemistry (e.g.\ reaction rates, electron-grain sticking coefficients; \citealt{Nishi1991,Bai2011}). These uncertainties could naturally affect models that include microphysics and lead to different outcomes when different ingredients are employed in, for example, the chemistry \citep[e.g.][]{Egan1996,Ilgner2006,Marchand2016,Wurster2016,Dzyurkevich2017}, the dust content \citep[e.g.][]{Nishi1991,Okuzumi2009,Ivlev2016,Zhao2016}, or cosmic rays \citep{Padovani2013}.

Our goal in this study is to determine which physical parameters are relevant for the evolution of the gas when self-consistently evolving time-dependent MHD alongside microphysics and chemistry \citep[e.g.][]{Kunz2009,Xu2016}, rather than by post-processing simulation snapshots \citep[e.g.][]{Padovani2013,Dzyurkevich2017}, or employing pre-computed equilibrium tables \citep[e.g.][]{Gressel2011,Marchand2016}. For this reason, we adopt a well-established, relatively simple framework for our experiments, i.e., a time-dependent, oblique, magnetic shock wave set in an environment that resembles the conditions of a pre-stellar core/dense molecular cloud \citep{Lesaffre2004,Chen2012,Hollenbach2013,Flower2015,Holdship2017,Nesterenok2018}. In this particular setting, the dominant non-ideal effect is ambipolar diffusion \citep{Draine1980,Smith2003,Duffin2008}.

There have been several other studies of non-ideal MHD shocks in dusty plasmas using both steady-state \citep[e.g.][]{Pilipp1994,Wardle1998,Chapman2006} and time-dependent approaches \citep[e.g.][]{vanLoo2009,Ashmore2010,vanLoo2013} that examine different microphysical effects. Expanding upon these works, this paper aims to compare the importance of several microphysical ingredients and, in particular, what role cosmic rays play in determining the structure and evolution of MHD shocks.

We have developed, applied, and made publicly available a 1D, non-ideal MHD code and pre-processor\footnote{\linktocode} to explore how common, simplifying assumptions about the microphysics affect the results relative to a full treatment of the problem. By varying the ingredients included in the experiments, we also identify simplifications to the chemistry and microphysics that \emph{do not} strongly affect the results and are therefore worth exploring for possible use in large-scale, multi-dimensional, non-ideal MHD simulations.

Our self-consistent, albeit simplified, formulation of the problem also makes it possible to identify feedback processes in the chemistry/microphysics which are responsible for non-linear responses to variations of the external or internal parameters. For instance, as discussed in \sect{sect:shock_parameters}, the indirect effect of cosmic rays on the gas temperature via ambipolar diffusion heating.

The paper is structured as follows: First, in \sect{sect:MHD1D} we introduce the equations of non-ideal MHD and describe their implementation. In Sects.\ \ref{sect:cooling} to \ref{sect:niMHD}, we discuss the details and assumptions made for cooling and heating, chemistry, cosmic rays, and non-ideal microphysics. We verify the results produced by the code (described in \sect{sect:code}) with a set of well-established tests in \sect{sect:benchmark} before investigating how varying the microphysical ingredients affect the results in \sect{sect:shock_parameters}. We conclude in \sect{sect:conclusions}.

\section{Methods: Non-ideal MHD 1D code}\label{sect:MHD1D}
To test the effects of microphysics in a physically motivated, non-linear and time-evolving environment, we developed a 1D, time-implicit MHD code. The code evolves the physical quantities, $\mathbf{U}$, forward in time via
\begin{equation}\label{eqn:FUSsys}
	\frac{\partial \mathbf{U}}{\partial t} = -\frac{\partial \mathbf{F}(\mathbf{U})}{\partial x} + \mathbf{S}(\mathbf{U})\,,
\end{equation}
where $\mathbf{F}$ are the fluxes, $\mathbf{S}$ the sources and sinks, and both are functions of $\mathbf{U}$. We define $\mathbf{U}=(\rho, \rho v_x, \rho v_y, \rho v_z, B_x, B_y, B_z, E, \rho X_i)$, where $\rho$ is the mass density, $v_i$ and $B_i$ are the \ith component of velocity and magnetic field, respectively, $E$ is the total energy density, and $X_i$ are the mass fractions of chemical species. $\mathbf{U}$ is defined at the centre of each cell.

Assuming that the ions and neutrals can be represented by a single fluid\footnote{cf.\ the more complex and numerically challenging multiple fluid approach \citep[e.g.][]{Ciolek2002,Falle2003,OSullivan2006}.} \citep{Shu1987,Choi2009,Masson2012}, \eqn{eqn:FUSsys} can be explicitly written, including ambipolar diffusion terms, as\footnote{We employ Gaussian cgs\ units, i.e., the permeability $\mu_0=1$.}:
\begin{eqnarray}\label{eqn:hydro_full}
	\partial_t \rho & = & - \partial_x \left[ \rho v_x \right],\label{eqn:hydro_full_first}\\
	\partial_t \left[ \rho v_x \right] & = & - \partial_x \left[ \rho v_x^2 + P^* - \frac{B_x^2}{4\pi} \right],\\
	\partial_t \left[ \rho v_y \right] & = & - \partial_x \left[ \rho v_xv_y - \frac{B_x B_y}{4\pi} \right],\\
	\partial_t \left[ \rho v_z \right] & = & - \partial_x \left[ \rho v_xv_z - \frac{B_x B_z}{4\pi} \right],\\
	\partial_t B_x & = & 0,\\
	\partial_t B_y & = & - \partial_x \left[ v_xB_y-v_yB_x \right.\nonumber\\
			&  & + \left. \frac{\eta_{\rm AD}}{\mathbf{B}^2} \left( F_{B,x}B_y - F_{B,y}B_x \right) \right],\label{eqn:by_evolution}\\
	\partial_t B_z & = & - \partial_x \left[ v_x B_z - v_zB_x \right.\nonumber\\\
			&  & + \left. \frac{\eta_{\rm AD}}{\mathbf{B}^2} \left( F_{B,z}B_x - F_{B,x}B_z \right) \right],\label{eqn:bz_evolution}\\
	\partial_t E & = & - \partial_x \left\{ \left(E + P^* \right) v_x - \frac{B_x}{4\pi} (\mathbf{v}\cdot \mathbf{B}) \right.\nonumber\\\label{eqn:hydro_full_energy}
			&  & - \frac{\eta_{\rm AD}}{4\pi \mathbf{B}^2} \left[ \left(F_{B,z}B_x - F_{B,x}B_z\right) B_z \right.\nonumber\\
			&  & - \left.\left. \left(F_{B,x}B_y - F_{B,y}B_x\right) B_y  \right] \right\}\nonumber\\
            &  & - \Lambda_{\rm chem} + \Gamma_{\rm CR},\\
	\partial_t \zeta & = & \mathcal{Z}(\rho, B_x, B_y, B_z),\label{eqn:cr_evolution}\\
	\partial_t \left[\rho X_i \right] & = & - \partial_x \left[ \rho X_i v_x \right] + \mathcal{P}_i - \rho X_i\mathcal{L}_i\,,\label{eqn:hydro_full_last}\label{eqn:hydro_full_chemistry}
\end{eqnarray}
where $\mathbf{B}$ is the modulus of the magnetic field\footnote{$\mathbf{B}^2={B_x^2+B_y^2+B_z^2}$}, and where $\partial_t \equiv \partial /\partial t$ and $\partial_x \equiv \partial /\partial x$ . Each chemical species is advected and their chemistry evolved according to \eqn{eqn:hydro_full_chemistry}; production ($\mathcal{P}_i$) and loss ($\mathcal{L}_i$) terms are discussed in \sect{sect:chemistry}.

The total pressure is
\begin{equation}
	P^* = P + \frac{\mathbf{B}^2}{8\pi}\,,
\end{equation}
while we assume an ideal equation of state for the thermal pressure
\begin{equation}\label{eqn:pressure}
	P = (\gamma-1) \left( E - \frac{\rho \mathbf{v}^2}{2} - \frac{\mathbf{B}^2}{8\pi} \right)\,,
\end{equation}
and related to the temperature $T$, needed by the chemistry, via the ideal gas law
\begin{equation}\label{eqn:pressure_temperature}
	P = \frac{\rho\,k_{\rm B}}{\mu m_{\rm p}} T\,,
\end{equation}
where $k_{\rm B}$ is Boltzmann's constant, $\gamma$ the adiabatic index, $\mu$ the mean molecular weight, and $m_{\rm p}$ the mass of the proton.
The Lorentz force components are
\begin{eqnarray}
	F_{B,x} & = & - B_y \cdot \partial_xB_y - B_z \cdot \partial_xB_z,\\
	F_{B,y} & = & B_y \cdot \partial_xB_x,\\
	F_{B,z} & = & B_z \cdot \partial_xB_x.
\end{eqnarray}
Spatial derivatives are evaluated using 2nd-order finite differences.

The ambipolar diffusion resistivity is given by
\begin{equation}\label{eqn:etaAD}
	\eta_{\rm AD} = c^2\left(\frac{\sigma_{\rm P}}{\sigma_{\rm P}^2 + \sigma_{\rm H}^2} - \frac{1}{\sigma_\parallel}\right)\,,
\end{equation}
where  $c$ is the speed of light, $\sigma_\parallel$, $\sigma_{\rm P}$, and $\sigma_{\rm H}$ are the parallel, Pedersen and Hall conductivities, respectively, and will be discussed in \sect{sect:niMHD}.

The temporal evolution of the cosmic ray ionisation rate in each cell, $\zeta$, is calculated using \eqn{eqn:cr_evolution}, and discussed in detail in \sect{sect:cosmic_rays}. Cosmic ray heating ($\Gamma_{\rm CR}$), as well as chemical cooling ($\Lambda_{\rm chem}$), are described in \sect{sect:cooling}.

In \eqnrange{eqn:hydro_full_first}{eqn:hydro_full_last},  since $\partial_y = \partial_z = 0$,  the solenoidal condition ($\nabla \cdot \mathbf{B}=0$) therefore requires that $\partial_x B_x=0$, which is guaranteed in the code by construction.

\begin{figure}
   \centering
       \includegraphics[width=.48\textwidth]{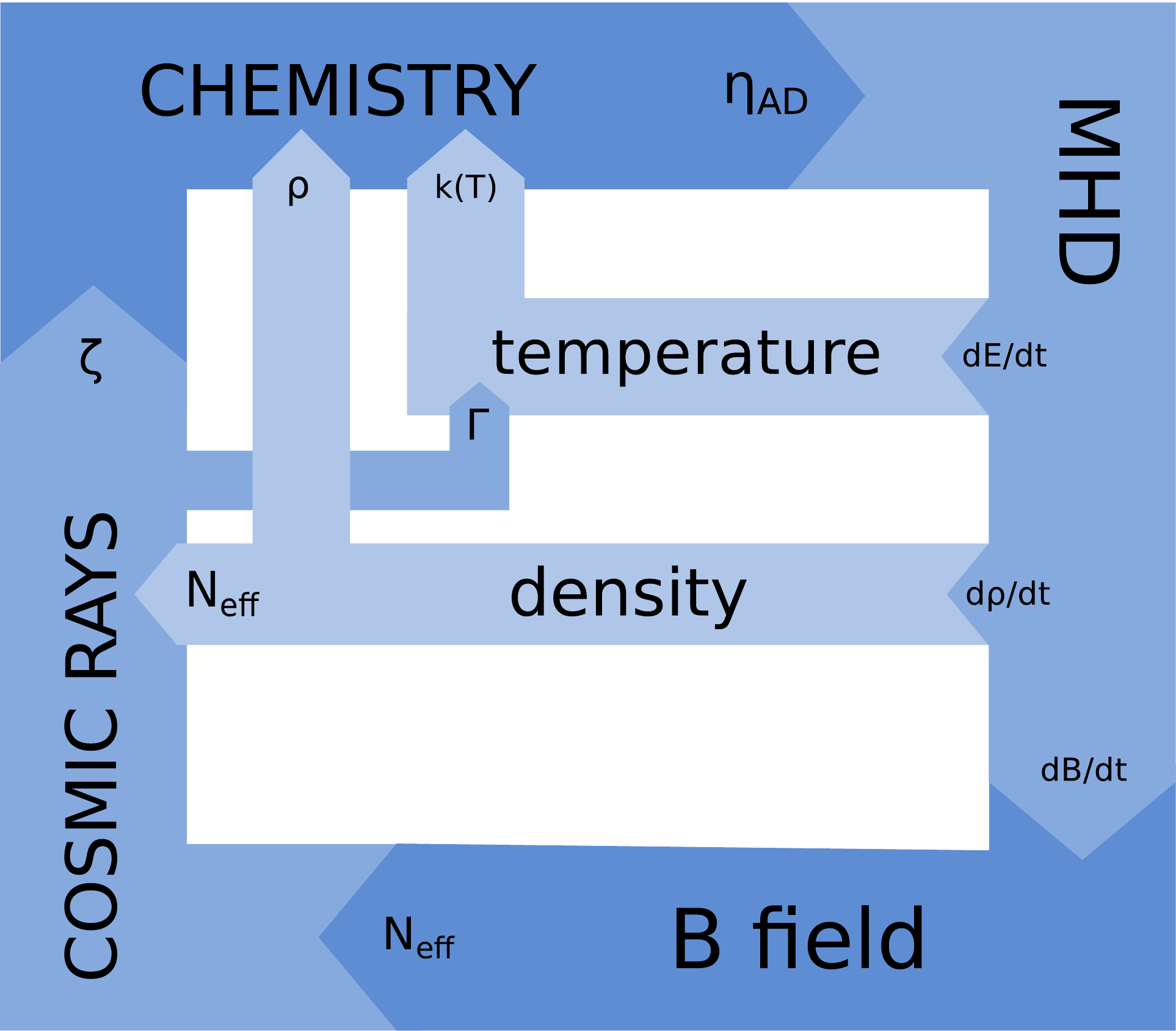}
 \caption{Sketch of the main processes included in our model; see \eqnrange{eqn:hydro_full_first}{eqn:hydro_full_last} and the text for further details.}
 \label{fig:scheme_all}
\end{figure}

The complex, non-linear interplay between the myriad of different processes described by \eqnrange{eqn:hydro_full_first}{eqn:hydro_full_last} is summarised in \fig{fig:scheme_all}: chemistry affects MHD via the resistivity coefficients ($\eta_{\rm AD}$), while MHD affects the energy ($E$), density ($\rho$), and magnetic field ($B$) evolution. Magnetic field and density determine the effective column density seen by cosmic rays ($N_{\rm eff}$), while chemistry depends on temperature via the reaction rate coefficients $k(T)$, on density, and on the cosmic ray ionisation rate ($\zeta$). Cosmic rays also affect temperature via direct heating ($\Gamma$).

\subsection{HLL solver}
We linearise the spatial derivatives on the right-hand side (RHS) of \eqn{eqn:FUSsys} using a standard HLL method \citep{Harten1997} in order to numerically calculate the fluxes. The flux in the \ith cell is given by
\begin{equation}
	\mathbf{F}_{i} = \frac{\mathbf{F}_{i+1/2}-\mathbf{F}_{i-1/2}}{\Delta x}\,,
\end{equation}
where $i\pm1/2$ denotes the quantity evaluated at the cell's interface. These are defined as
\begin{equation}
	\mathbf{F}_{i+1/2} = \frac{\alpha^+\mathbf{F}_i + \alpha^-\mathbf{F}_{i+1} - \alpha^+\alpha^-\left(\mathbf{U}_{i+1}-\mathbf{U}_{i}\right)}{\alpha^+ + \alpha^-}\,,
\end{equation}
where
\begin{equation}
	\alpha^\pm=\max\left|\pm\lambda^\pm_i, \pm\lambda^\pm_{i+1},0 \right|\,,
\end{equation}
$\lambda^\pm_i = v_x\pm c_{\rm f}$ are the eigenvalues of the Riemann problem at the cell interfaces, and the fast magnetosonic velocity evaluated at $i$ is
\begin{equation}
	c_{\rm f}^2=\frac{1}{2}\left(\theta+\sqrt{\theta^2-4\,c_{\rm s}^2\frac{B_x^2}{4\pi\rho}}\right)\,,
\end{equation}
with $\theta=v_{\rm A}^2+c_{\rm s}^2$, the speed of sound $c_{\rm s}^2=\gamma P/\rho$, and the Alfv\'en speed $v_{\rm A}^2=\mathbf{B}^2/(4\pi\rho)$.

\subsection{Implicit time integration}
We employ the \textsc{DLSODES} solver \citep{Hindmarsh1983, Hindmarsh2005} to integrate the system (\eqnrange{eqn:hydro_full_first}{eqn:hydro_full_last}) forward in time. This approach avoids the need to explicitly define a time-step using a standard Courant condition; only absolute and relative tolerances of the individual quantities are needed (see below). The \textsc{DLSODES} solver has access to the RHS of \eqn{eqn:FUSsys} \emph{for all grid points} and, for the sake of simplicity, we use the internally-generated Jacobian\footnote{For additional details, refer to the solver documentation contained in the \file{opkdmain.f} file.}. Our approach is fully implicit in time and does not require any operator splitting to solve, e.g., the chemistry or cooling alongside the MHD. The code has been successfully validated against a set of standard numerical experiments which are discussed in \appx{sect:tests}.

The accuracy of the method is determined by absolute ($\varepsilon_{\rm atol}$) and relative ($\varepsilon_{\rm rtol}$) tolerances\footnote{Tolerances are employed by the solver to compute the local error associated with the quantity $y$ as $\varepsilon_{\rm loc}=\varepsilon_{\rm rtol}|y| + \varepsilon_{\rm atol}$. Smaller tolerances increase the accuracy of the calculation, but could considerably increase the computational time.} defined for each variable and each cell. We set\footnote{Units of $\varepsilon_{\rm atol}$ and $\varepsilon_{\rm rtol}$ are in code units, that assumes cgs.} $\varepsilon_{\rm rtol}=10^{-8}$ for all variables, while $\varepsilon_{\rm atol}=10^{-30}$ for density ($\rho$), $\varepsilon_{\rm atol}=10^{-25}$ for the momentum ($\rho v_i$) and energy ($E$), $\varepsilon_{\rm atol}=10^{-10}$ for the magnetic field ($B_i$), and  $\varepsilon_{\rm atol}=10^{-20}$ for the cosmic ray ionisation rate ($\zeta$). The chemistry (i.e.\ $\rho X_i$), meanwhile, uses $\varepsilon_{\rm atol}=10^{-30}$ for all species. In principle, the solver allows for different tolerances in different cells, but we find this unnecessary in the current study.

\section{Methods: Cooling and Heating}\label{sect:cooling}
In addition to the usual ambipolar MHD heating and cooling processes stated explicitly in \eqn{eqn:hydro_full_energy}, we also include a simplified radiative chemical cooling and direct heating from cosmic rays. The former is taken from \theireqn{11} of \citet{Smith1997}:
\begin{equation}
  \label{eqn:chemical_cooling}
 \Lambda_{\rm chem} = 4.2\times10^{-31}\, n({\rm H_2})\, T^{3.3}\,{\rm erg~cm^{-3}~s^{-1}}\,,
\end{equation}
where $n({\rm H_2})$ is the molecular hydrogen number density in cm$^{-3}$, and $T$ is the gas temperature in K. The cooling function employed here is accurate enough given the chemistry model we adopt for the current investigation (see \sect{sect:chemistry}). In \sect{sect:shock_parameters}, however, we test the effects of varying the strength of the cooling term.

We model the cosmic ray heating as
\begin{equation}
  \label{eqn:cr_heating}
 \Gamma_{\rm CR} = \zeta\,Q(n_{\rm H})\,n({\rm H_2}) \,{\rm erg~cm^{-3}~s^{-1}}\,,
\end{equation}
where $\zeta$ is the cosmic ray ionisation rate (see \sect{sect:cosmic_rays}), $n_{\rm H}$ is the number density of H~nuclei ($n_{\rm H}=2n_{\rm H_2}$ in our chemical network), and $Q(n_{\rm H})$ is the heat deposited in the gas per
ionisation event, taken from Fig.\ 2 of \citet{Galli2015} \citep[see also][]{Glassgold2012} and fit here using
\begin{equation}\label{eqn:fit_heating_cr}
 Q(n_{\rm H}) = \sum_{i=0}^5 c_i \log\left(n_{\rm H}\right)^i\,{\rm eV}\,,
\end{equation}
using the coefficients found in \tab{tab:fit_heating_cr}. The fitting function is valid in the range $n_{\rm H}=10^2$ to $10^{10}$~cm$^{-3}$.

\begin{table}
\centering
\begin{tabular}{llllll}
\hline
 $i$ & $c_i$ & $i$ & $c_i$ & $i$ & $c_i$\\
\hline
   0 &  6.882876 &    2 &  -0.532834 &   4 &  -0.016907 \\
   1 &  2.231421 &    3 &   0.146966 &   5 &  0.000642 \\
\hline
\end{tabular}\caption{Coefficients used in \eqn{eqn:fit_heating_cr} to fit the amount of heat deposited per cosmic ray ionisation following. Note that units are eV.
}
\label{tab:fit_heating_cr}
\end{table}

\section{Methods: chemistry}\label{sect:chemistry}

To maintain a reasonable level of control over the many parameters in our model, we employ a simplified chemical network that follows the approach of \cite{Fujii2011} (see their Fig.\ 2 and our \tab{table:chemistry}). Despite this reduced model, in \sect{sect:benchmark}, we demonstrate that this network is capable of reproducing the results of a few more complicated chemical networks. Our model assumes that the ionisation of H$_2$ produces a cascade of fast reactions that lead immediately to e$^-$ and Mg$^+$ as products, where the latter is a proxy for all cations. Analogously, Mg$^+$ quickly recombines with electrons (and negatively charged grains) to reform H$_2$. In our model, the molecular hydrogen ionisation rate coefficient $k_{\rm H_2}$ is equal to the cosmic-ray ionisation rate $\zeta$ (see \sect{sect:cosmic_rays}). The rate of Mg$^+$ recombination is obtained from \citet{verner1996} in the form\footnote{\url{http://www.pa.uky.edu/~verner/rec.html}}:
\begin{equation}\label{eqn:gas_recombination}
k_{\rm rec}(T)= k_0 \left[ \sqrt{\frac{T}{T_0}} \left( 1 + \sqrt{\frac{T}{T_0}}\right)^{1-b} \left( 1 + \sqrt{\frac{T}{T_1}}\right)^{1+b} \right]^{-1}\,,
\end{equation}
with $k_0 = 1.92\times10^{-11}$~cm$^{3}$~s$^{-1}$, $b=0.3028$, $T_0=4.849\times10^{2}$~K, and $T_1=5.89\times10^{6}$~K.
Since Mg$^+$ is a proxy for all positive ions, $k_{\rm rec}$ is therefore an effective recombination rate. In \sect{sect:shock_parameters}, we will examine the effects of varying $k_{\rm rec}$.

\subsection{Differential equations for chemistry}\label{sect:ode_chemistry}
Differential equations for the production rate $\mathcal{P}_i$ and loss rate $\rho X_i\mathcal{L}_i$ of the \ith species are solved simultaneously with the equations of MHD in a single system, and are defined\footnote{\eqn{eqn:chemistry_production} and \eqn{eqn:chemistry_loss} represent a standard set of differential equations for chemistry, but where the species abundances are given by their mass density instead of the more typical number density.} by
\begin{eqnarray}
 \mathcal{P}_i &=& m_i \rho^2 \sum_{r_1,r_2} k_{r_1,r_2} \frac{X_{r_1}}{m_{r_1}} \frac{X_{r_2}}{m_{r_2}},\label{eqn:chemistry_production}\\
 X_i\rho\mathcal{L}_i &=& X_i\rho^2 \sum_{r_1} k_{r_1} \frac{X_{r_1}}{m_{r_1}}\,, \label{eqn:chemistry_loss}
\end{eqnarray}
where $k_{r_1,r_2}$ is the reaction rate coefficient between species $r_1$ and $r_2$, while $m_i$ and $X_i$ are the mass and the mass fraction of the \ith species, such that $n_im_i=\rho X_i$.

Together with the chemical network (defined in the previous Section), Eqs.\ (\ref{eqn:chemistry_production}) and (\ref{eqn:chemistry_loss}) conserve the total number density, but not the total mass, because an H$_2$ molecule is instantaneously converted into an Mg$^+$ atom that is 24 times more massive. In principle, for the standalone chemical network, this does not represent a problem because the number density is conserved by construction. However, the hydrodynamics advects the mass density of the species, and it is therefore crucial to ensure conservation of mass. To avoid this issue, we define the mass of Mg$^+$ as $m_{\rm Mg^+}'=m_{\rm H_2} - m_{\rm e^-}$. When using the actual mass of Mg$^+$, we find that the relative error on total mass conservation can be as large as $10^{-4}$ (instead of $\lesssim 10^{-7}$), while the error on global charge can reach $10^{-2}$ in the worst cases (instead of $\lesssim 10^{-8}$). We therefore use $m_{\rm Mg^+}'$ for our models. We also note that a non-reduced chemical network will not, in general, be affected by this problem, since the mass will be correctly conserved by each reaction.

\begin{table}
\begin{tabular}{llllll}
\hline
 H$_2$ 	&		& $\to$ & Mg$^+$ + e$^-$ 	& $k_{\rm H_2}$ & $k_{\rm H_2}=\zeta$\\
 Mg$^+$ & + e$^-$ 	& $\to$ & H$_2$ 		& $k_{\rm rec}$ 	& \eqn{eqn:gas_recombination}\\
 Mg$^+$ & + g$(Z>0)$	& $\to$ & H$_2$ + g$(Z+1)$ 	& $k_{i,j}^+$ 	& \eqn{eqn:dust_k_repulse}\\
 Mg$^+$ & + g$^0$	& $\to$ & H$_2$ + g$^+$ 	& $k_{i,j}^0$ 	& \eqn{eqn:dust_k_neutral}\\
 Mg$^+$ & + g$(Z<0)$ 	& $\to$ & H$_2$ + g$(Z+1)$ 	& $k_{i,j}^-$ 	& \eqn{eqn:dust_k_attract}\\
 e$^-$  & + g$(Z>0)$	& $\to$ & g$(Z-1)$ 		& $k_{i,j}^-$	& \eqn{eqn:dust_k_attract}\\
 e$^-$  & + g$^0$	& $\to$ & g$^-$ 		& $k_{i,j}^0$	& \eqn{eqn:dust_k_neutral}\\
 e$^-$  & + g$(Z<0)$ 	& $\to$ & g$(Z-1)$ 		& $k_{i,j}^+$ 	& \eqn{eqn:dust_k_repulse}\\
 g$^-$  & + g$^+$ 	& $\to$ & g$^0$ + g$^0$		& $k_{i,j}^-$ 	& \eqn{eqn:dust_k_attract}\\
 g$^{--}$ & + g$^{++}$	& $\to$ & g$^0$ + g$^0$		& $k_{i,j}^-$ 	& \eqn{eqn:dust_k_attract}\\
 g$^{++}$ & + g$^{0}$	& $\to$ & g$^+$ + g$^+$		& $k_{i,j}^0$ 	& \eqn{eqn:dust_k_neutral}\\
 g$^{--}$ & + g$^{0}$	& $\to$ & g$^-$ + g$^-$		& $k_{i,j}^0$ 	& \eqn{eqn:dust_k_neutral}\\
 g$^{++}$ & + g$^-$	& $\to$ & g$^0$ + g$^+$		& $k_{i,j}^-$ 	& \eqn{eqn:dust_k_attract}\\
 g$^{--}$ & + g$^+$	& $\to$ & g$^0$ + g$^-$		& $k_{i,j}^-$ 	& \eqn{eqn:dust_k_attract}\\
\hline
\end{tabular}\caption{List of reactions in our reduced network, the rate coefficient symbol, and reference to the text. Symbols g$(Z<0)$ and g$(Z>0)$ indicate grains with negative and positive charge, respectively, while g$(Z+1)$ and g$(Z-1)$ indicate the reactant grain plus or minus one charge.}
\label{table:chemistry}
\end{table}

\subsection{Grain chemistry}\label{sect:grain_chemistry}
In order to determine the fraction of charged species to compute the resistivity coefficients (see \sect{sect:nimhd}), we include dust grains that can recombine with electrons and exchange charge with cations and amongst themselves (see \tab{table:chemistry}). We integrate the grain size distribution $\varphi(a)$ over size range $a_{\rm min}$ to $a_{\rm max}$ to provide averaged reaction rate coefficients, $k(a, T)$, that are functions of the grain size $a$:
\begin{equation}\label{eqn:average_k_dust}
 \langle k(T)\rangle = \frac{\int_{a_{\rm min}}^{a_{\rm max}} \varphi(a) k(a,T) \dd a}{\int_{a_{\rm min}}^{a_{\rm max}} \varphi(a) \dd a}
\end{equation}
for particle-grain interactions, and
\begin{equation}\label{eqn:average_k_dust_dust}
 \langle k(T)\rangle = \frac{\int_{a_{\rm min}}^{a_{\rm max}}\int_{a_{\rm min}}^{a_{\rm max}} \varphi(a) k(a,a',T) \dd a\, \varphi(a') \dd a'}{\int_{a_{\rm min}}^{a_{\rm max}}\int_{a_{\rm min}}^{a_{\rm max}} \varphi(a) \dd a\, \varphi(a') \dd a'}
\end{equation}
for grain-grain interactions, where $k(a, a', T)$ are the rate coefficients for collisions of grains with radius $a$ and $a'$, respectively.

Following \citet{Draine1986}, for reactants with opposite charge ($Z_iZ_j<0$, e.g.\ electrons and positively charged grains), we include the Coulomb factor and the charge focusing due to polarisation as
\begin{eqnarray}\label{eqn:dust_k_attract}
 k_{i,j}^-(a_s, T) & = & \pi v_{\rm g} a_s^2\left( 1 - \frac{Z_i Z_j q^2}{a_s k_{\rm B} T} \right)\nonumber\\
	& \cdot & \left[1+\sqrt{\frac{2q^2Z_i^2}{a_s k_B T - 2 Z_i Z_j q^2}}\right] \,S(T)\,,
\end{eqnarray}
where  $q$ is the elemental charge, $Z_i$ is the charge of the particle (e.g.\ electrons have $Z_i=-1$), $Z_j$ is the charge of the grain\footnote{When two grains interact, $Z_i$ and $Z_j$ are the charge counts of the grains. Since \citet{Draine1986} consider only an interaction between a conducting sphere and a test charge (see their sect.\ II.a), we assume that $j$ is always the larger collision partner, i.e., a grain in the grain-particle collision, and the smaller grain in grain-grain interactions.}, $a_s=a_i+a_j$ is the sum of the grain sizes that reduces to $a_s=a_j$ when $i$ is a particle, $S(T)$ is the sticking coefficient (to be discussed below), and
\begin{equation}
 v_g = \sqrt{\frac{8 k_{\rm B} T}{\mu_{i,j} m_p}}\,,
\end{equation}
is the gas thermal velocity, $\mu_{i,j} = m_im_j / (m_i+m_j)$ is the reduced mass of the two species, and $m_{\rm p}$ the proton mass.

Analogously, for reactants with repulsive charges ($Z_iZ_j> 0$, e.g., a cation and a positively charged grain), the rate is
\begin{eqnarray}\label{eqn:dust_k_repulse}
 k_{i,j}^+(a_s, T) &=& \pi v_{\rm g} a_s^2 \left[ 1 + \left(\frac{4 a_s k_{\rm B} T}{q^2 Z_i^2} + 3\frac{Z_j}{Z_i}\right)^{-1/2} \right]^2\nonumber\\
 & \cdot & \exp\left(-\frac{\theta_v q^2 Z_i^2}{a k_{\rm B} T}\right)\,S(T)\,,
\end{eqnarray}
with $\theta_v=Z_j^{3/2}/[Z_i(\sqrt{Z_i}+\sqrt{Z_j})]$.

Finally, for $Z_iZ_j=0$ (e.g.~a cation and a neutral grain), we have
\begin{equation}\label{eqn:dust_k_neutral}
 k_{i,j}^0(a_s, T) = \pi v_{\rm g} a_s^2 \left(1+\sqrt{\frac{\pi q^2 Z_i^2}{2 a_s k_{\rm B} T}}\right)\,S(T)\,.
\end{equation}

\begin{figure}
   \centering
       \includegraphics[width=.49\textwidth]{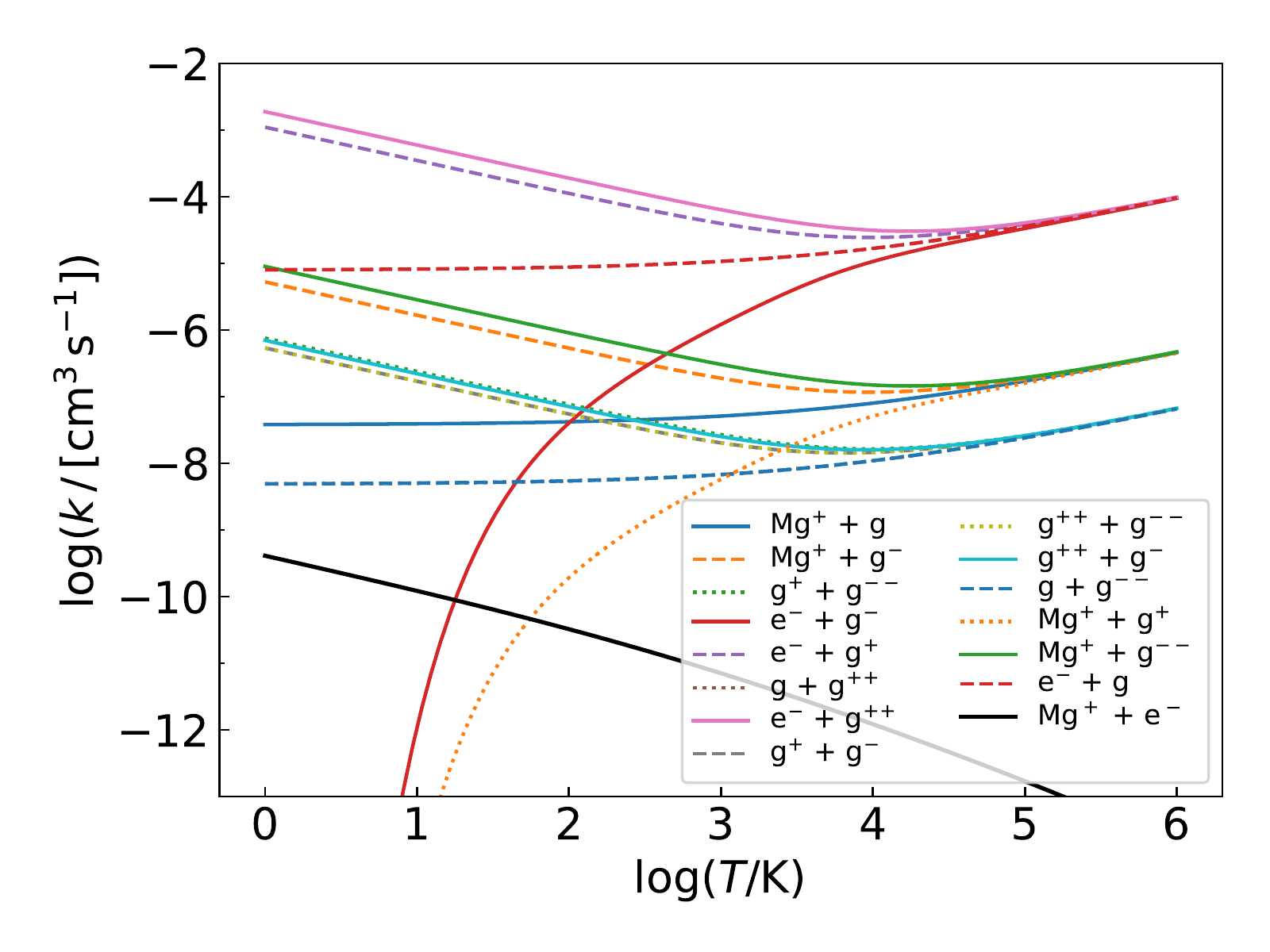}
 \caption{Grain chemistry rate coefficients as a function of temperature. Products are omitted from the legend for clarity. As expected, electron-positive grain rate coefficients reach larger values for temperatures below $10^4$~K, while rate coefficients for repulsive reactants quickly drop as the temperature decreases. In this plot, we assume a power-law distribution for the grain size distribution, $\varphi(a)\propto a^p$ with $p=-3.5$, $a_{\rm min}=10^{-7}$ to $a_{\rm max}=10^{-5}$~cm. For the sake of comparison, we include the rate coefficient for the recombination of Mg$^+$ with electrons (i.e.\ $k_{\rm rec}$).}
 \label{fig:grain_rates}
\end{figure}

For grain-particle interactions, we compute $\langle k_{i,j}^{\pm,0}(T) \rangle$ by using \eqn{eqn:average_k_dust}  together with \eqn{eqn:dust_k_attract}, \eqn{eqn:dust_k_repulse}, or \eqn{eqn:dust_k_neutral}, and assuming $a_s=a_j$. Grain-grain interactions are modelled analogously but using $a_s=a_i+a_j$ and \eqn{eqn:average_k_dust_dust} instead of \eqn{eqn:average_k_dust}. All of the rate coefficients discussed in this Section are shown for comparison in \fig{fig:grain_rates}.

To speed-up code execution, we pre-compute the grain rate coefficients as a function of the temperature and apply a linear fitting function in logarithmic space at run-time. The grain size distribution properties are discussed in \appx{sect:grain_distribution}, while fitting functions for reactions involving grains are given in \appx{sect:k_poly_grains}.

\subsection{Sticking coefficient}\label{sect:sticking}
To model the electron-grain sticking coefficient, $S(T)$, in \eqn{eqn:dust_k_attract} we refer to the appendix of \citet{Nishi1991} as well as \citet{Bai2011}. Electrons, because of their excess energy, only stick with some probability when they encounter a grain. A key parameter is $D$, the depth of the potential well between electrons and grains due to the polarisation interaction. We do not implement the equations reported there, but we fit Fig.\ 6 in  \citet{Bai2011} assuming $D=1$~eV and $a=0.1$~$\mu$m. The fit is $\log[S(T, Z)] = c_{Z,1} \log(T)^2 + c_{Z,2} \log(T) + c_{Z,3}$, where $Z$ is the grain charge and the coefficients are listed in \tab{table:fit_stick}. Using $a=0.1$~$\mu$m (instead of integrating over the grain size distribution) may lead to errors, but since the choice of $D$ is arbitrary, we consider this fit accurate enough for the aims of this work. Moreover, comparing the two panels from Fig.~6 of \citet{Bai2011}, we note that the sticking coefficient with neutral grains (i.e.\ the main electron sticking route in our model) is almost size-independent. Nevertheless, in \sect{sect:shock_parameters}, we will vary the sticking coefficient and demonstrate its impact on the results.

As for cation-grain and grain-grain sticking coefficients, following \citet{Draine1986}, we set $S(T)=1$ for the interaction between positive ions and any type of grain; in \eqn{eqn:dust_k_attract}, \eqn{eqn:dust_k_repulse}, and \eqn{eqn:dust_k_neutral}, $S(T)<1$ only when the grain partner is an electron.

\begin{table}
\begin{tabular}{rrrr}
\hline
   $Z$ & $c_{Z,1}$ &  $c_{Z,2}$ &  $c_{Z,3}$ \\
\hline
   $-4$ &  -0.41953296 &  0.37771378 &   0.06950703 \\
    $-3$ & -0.41819111 &  0.34261771 &   0.18453260 \\
    $-2$ & -0.40908288 &  0.30922328 &   0.20339583 \\
    $-1$ & -0.39206456 &  0.26730250 &   0.16134104 \\
    $0$ &  -0.36848365 &  0.22498770 &   0.05100280 \\
    $1$ &  -0.34350388 &  0.21428663 &  -0.17200160 \\
    $2$ &  -0.33385135 &  0.31951247 &  -0.60550265 \\
    $3$ &  -0.35192170 &  0.55369825 &  -1.17743478 \\
    $4$ &  -0.28404485 &  0.18312184 &  -0.70825975 \\
\hline
\end{tabular}\caption{Coefficient for the electron sticking $S(T, Z)$. See text for further details and its implementation in the \file{rates.f90} file. }
\label{table:fit_stick}
\end{table}

\section{Methods: cosmic rays}\label{sect:cosmic_rays}
Being charged particles, cosmic rays follow helical trajectories around magnetic field lines as they propagate. As a consequence, in the presence of a magnetic field, the effective column density $N_{\rm eff}$ ``seen'' by a cosmic ray can be much larger than the line of sight column density, especially if the magnetic field is not laminar \citep{Padovani2013}.

We consistently compute the propagation of cosmic rays following the approach of \citet{Padovani2018}, where the cosmic ray ionisation rate of H$_2$, $k_{\rm H_2}=\zeta$, is a function of the effective column density travelled by the particle, $\zeta\equiv f(N_{\rm eff})$, and is described in Appendix F of \citet{Padovani2018}. Since we consider that our shock wave occurs in the vicinity of a pre-stellar core, prior to the cosmic rays entering our simulation domain, we assume they are partially attenuated by the surrounding medium. Thus, following \citet{Ivlev2015}, we assume that the initial effective column density experienced by the cosmic rays is $N_{{\rm eff},1}=5\times10^{21}$~cm$^{-2}$. This value is evaluated at the centre of the first cell (thus, the ``$1$'' subscript). Since cosmic rays gyrate around magnetic field lines, and given the periodicity of the simulation domain along $y$- and $z$-directions, we compute the effective distance travelled as
\begin{equation}\label{eqn:dxeff}
	\Delta x_{\rm eff} = \frac{\Delta x}{\cos \vartheta \cos \phi}\,,
\end{equation}
where $\vartheta = \arctan(B_z/B_x)$ and $\phi = \arctan(B_y / \sqrt{B_z^2+B_x^2})$, while $B_x$, $B_y$, and $B_z$ are evaluated at the cell interface. We then compute the effective column density at the centre of the \ith cell as
\begin{equation}\label{eqn:sum_neff}
	N_{{\rm eff},i} = N_{{\rm eff},1} + \sum_{j=2}^i \Delta x_{{\rm eff},j} n_j\,,
\end{equation}
where $\Delta x_{{\rm eff},j}$ and $n_j$ are calculated at the interface between cells $j-1$ and $j$ using a linear interpolation. Once the effective column density is available, we can retrieve the H$_2$ ionisation rate $\zeta_i$ at the centre of the \ith cell.

Within the column density range we are interested in here ($10^{20}<N<10^{23}$~cm$^{-2}$),
$\zeta_i$ is fairly represented by a power-law
\begin{equation}\label{eqn:cr_fit}
 \zeta_i = b_1\left( N_{{\rm eff},i} \right)^{b_2}\,;
\end{equation}
the coefficients $b_1$ and $b_2$ are discussed below.

Since $\cos{(\arctan{x})} = \left({x^2+1}\right)^{-1/2}$, using the definitions of $\vartheta$ and $\varphi$, we can rewrite \eqn{eqn:dxeff} as
\begin{equation}
 \Delta x_{{\rm eff},j} = \frac{\Delta x }{B_x}\mathbf{B}_j\,,
\end{equation}
and thus
\begin{equation}
 N_{{\rm eff},i} = N_{{\rm eff},1} + \frac{\Delta x }{B_x}\sum_{j=2}^i \mathbf{B}_j n_j\,.
\end{equation}
Substituting this into \eqn{eqn:cr_fit}, and assuming that $\Delta x$ and $B_x$ are constant in time, we then differentiate with respect to time and obtain
\begin{eqnarray}\label{eqn:dzeta_dt}
 \frac{\partial\zeta_i}{\partial t} & = &  \frac{b_1 b_2 \Delta x}{B_x} \left( N_{{\rm eff},1} + \frac{\Delta x}{B_x}\sum_{j=2}^i n_j \mathbf{B}_j \right)^{b_2-1}\\
	& \cdot & \sum_{j=2}^i\left[ \frac{\partial n_j }{\partial t} \mathbf{B_j} + \frac{n_j}{\mathbf{B}_j} \left( B_{y,j} \frac{\partial B_{y,j}}{\partial t} + B_{z,j} \frac{\partial B_{z,j}}{\partial t}  \right)   \right]\,.\nonumber
\end{eqnarray}

We note that the $\zeta_i$ in a given cell depends (non-trivially) on the densities and magnetic field values of all the cells from the first to the \ith, i.e., on $3\times i$ variables. In practice, this considerably reduces the internal time step of \dlsodes, since the number of variables that $\zeta_i$ depends on is large and the Jacobian becomes considerably less sparse. In fact, by using a constant $\zeta$, the integration time can be reduced by a factor of approximately one hundred.

In principle, when computing the propagation of cosmic rays, one should account for the effects of magnetic focusing and mirroring \citep{Cesarsky1978,Desch2004,Padovani2011}. Focusing and mirroring mechanisms act to amplify and reducing the cosmic ray flux, respectively, and could be important in regions of star formation. However, \citet{Silsbee2018} has demonstrated that these two effects nearly cancel each other out when the magnetic field strength has a single peak along the field lines, which is indeed the case in this work. Therefore, in the following, we choose to neglect mirroring and focusing effects.

The propagation of cosmic rays can also be affected by scattering due to self-generated Alfv\'en waves \citep{Skilling1976, Hartquist1978}, but this mechanism is only important at the edges and the more diffuse parts of a molecular cloud, and thus we can safely neglect it in this work.

\subsection{Lower and upper bounds of $\zeta_i$}\label{sect:cosmic_rays_LH}
The cosmic ray ionisation rate at a given column density $N$ is given by
\begin{equation}
 \zeta(N) = 4\pi\int j(E,N)[1+\Phi(E)]\sigma^{\rm ion}(E){\rm d}E\,,
\end{equation}
where $j(E)$ is the cosmic ray differential flux (also called the spectrum), $\Phi$ is a multiplicity factor accounting for ionisation by secondary electrons, and $\sigma^{\rm ion}$ is the ionisation cross section. Since $\sigma^{\rm ion}$ is known to peak at low energies, the maximum contribution to $\zeta$ comes from cosmic rays in the energy range $10~{\rm MeV}\lesssim E\lesssim 1~{\rm GeV}$ (\citealt{Padovani2009}).

The most recent Voyager~1 data release~\citep{Cummings2016} leads to the conclusion that no upturn is expected in the interstellar proton spectrum down to energies of at least 3~MeV. The corresponding ionisation rate, however, is more than a factor of 10 smaller than estimates from observations in diffuse clouds \citep{Indriolo2015,Neufeld2017}. For this reason, as in \citet{Padovani2018}, we consider two different models for the cosmic ray proton spectrum: a ``low'' spectrum, obtained by extrapolating the Voyager~1 data to low energies, and a ``high'' spectrum. The latter can be considered as an upper bound to the actual average galactic cosmic ray spectrum
and provides an upper limit to the values of $\zeta$ estimated for diffuse clouds. The resulting ionisation rates and their comparison to observations is discussed in \citet{Ivlev2015}.

The values of $b_{1}$ and $b_{2}$ in \eqn{eqn:cr_fit} that are needed to reproduce the two trends in $\zeta$ are $b_1=1.327\times10^{-12}$~s$^{-1}$ and $b_2=-0.211$ for the ``low'' case, and $b_1=5.34\times10^{-5}$~s$^{-1}$ and $b_2=-0.384$ for the ``high'' case. Note that the validity of the fit is limited to $10^{20} < N < 10^{23}$~cm$^{-2}$.

\section{Methods: Non-ideal MHD coefficients}\label{sect:niMHD}\label{sect:nimhd}
Ambipolar diffusion is controlled by the $\eta_{\rm AD}$ resistivity coefficient, appearing in \eqnrange{eqn:by_evolution}{eqn:hydro_full_energy} and defined in \eqn{eqn:etaAD}. The resistivity coefficient depends on conductivities (i.e.\ $\sigma_{\parallel}$, $\sigma_{\rm P}$, and $\sigma_{\rm H}$) that are functions of temperature, magnetic field, and species abundances. Given the reduced chemistry that we include in our model (\sect{sect:chemistry}), we assume that the only interactions that affect the conductivity are collisions between charged particles (electrons, cations, and grains) and molecular hydrogen. In principle, if we were to follow \citet{Pinto2008b}, each charged species could exchange momentum with any other species in the gas, but, since the momentum transfer is dominated by the interaction between charged species and H$_2$ (which in our model is the main neutral component of the gas), we do not explicitly include all interactions. For the sake of completeness, however, we report all the possible interactions from \citet{Pinto2008b} in \appx{sect:pinto_collisions}.

\subsection{Conductivities}\label{sect:beta}\label{sect:sigma}
The three conductivities (parallel, Pedersen, and Hall) are given by \citep[e.g.][]{Pinto2008a}:
\begin{eqnarray}\label{eqn:conductivities}
 \sigma_{\parallel} & = & \frac{c}{B} \sum_i \frac{q Z_i \rho_i}{m_i} \beta_{i,n},\\
 \sigma_{P} & = & \frac{c}{B} \sum_i \frac{q Z_i \rho_i}{m_i} \frac{\beta_{i,n}}{1+\beta_{i,n}^2},\\
 \sigma_{H} & = & \frac{c}{B} \sum_i \frac{q Z_i \rho_i}{m_i} \frac{1}{1+\beta_{i,n}^2}\,,
\end{eqnarray}
where $q$ is the elementary charge, $m_i$ is the mass of a charged particle, $qZ_i$ is its charge, $\rho_i$ its mass density, and $\beta_{i,n}$ is the Hall parameter which takes into account the interaction between charged particles and neutral species (in our case H$_2$). The sum is over electrons, cations, and charged dust grains.

The Hall parameter for collisions between the \ith charged particle (gas or dust) and a neutral species is given by
\begin{equation}
 \beta_{i,n} = \left(\frac{qZ_i B}{m_i c}\right) \frac{m_i + m_n}{\rho_n R_{i,n}(T)}\,,
\end{equation}
where $\rho_n$ is the neutral gas mass density, $m_n$ its mass, and $R_{i,n}(T)$ is the momentum exchange rate coefficient, which is described in the next Section.

\subsection{Momentum transfer rate coefficients}
\subsubsection{Charged grains--H$_2$}\label{sect:charged_grain_H2}
To model the interaction between charged dust grains and molecular hydrogen, we follow Sect.\ 6 of \citet{Pinto2008b}. We employ their \theireqn{25} when the condition in their \theireqn{23} is satisfied (the hard sphere approximation rate; $R_{\rm hs}$), otherwise we use their \theireqn{A.3} (the Langevin rate; $R_{\rm L}$). Assuming that, at a critical grain size $a_{\rm c}$, \theireqn{25} and \theireqn{A.3} are equal, we can write
\begin{equation}\label{eqn:alpha_crit}
 a_{\rm c}(T, Z) = \frac{0.206}{\sqrt{\delta}} \left( \frac{\alpha_{\rm pol} |Z|}{T} \right)^{1/4}\,,
\end{equation}
where $\delta=1.3$ is taken from \citet{Liu2003} and $\alpha_{\rm pol}=8.06\times10^{-25}$~cm$^3$ is the polarisability of molecular hydrogen \citep{Pinto2008b}.

Adopting a grain size distribution $\varphi(a)\propto a^p$ over size range $a_{\rm min}$ to $a_{\rm max}$ (\sect{sect:grain_chemistry}), and the Langevin rate $R_{L}$ is size-independent, the rate coefficient then becomes
\begin{eqnarray}\label{eqn:charge_grain_H2}
 \langle R_{g,n}(T,Z) \rangle & = &\frac{ R_{\rm L}\int_{a_{\rm min}}^{a_{\rm c}} \varphi(a)\,\dd a + \int_{a_{\rm c}}^{a_{\rm max}}R_{\rm hs}\, \varphi(a)\,\dd a}{\int_{a_{\rm min}}^{a_{\rm max}} \varphi(a)\,\dd a}\nonumber\\
  & = & 2.21\pi\sqrt{\frac{\alpha_{\rm pol}|Z|q^2}{m_{\rm H_2}}} \cdot\frac{a_{\rm c}^{p+1} - a_{\rm min}^{p+1}}{a_{\rm max}^{p+1} - a_{\rm min}^{p+1}}\nonumber\\
 & + & v_{\rm g}(T) \frac{4\pi\delta}{3}\cdot\frac{a_{\rm max}^{p+3} - a_{\rm c}^{p+3}}{a_{\rm max}^{p+1} - a_{\rm min}^{p+1}}\frac{p+1}{p+3}\,,
\end{eqnarray}
where $m_{\rm H_2}$ is the mass of molecular hydrogen and, since the mass of the grain is larger than the mass of H$_2$, the reduced mass is \mbox{$\mu\sim m_{\rm H2}$}. A simplified expression for \eqn{eqn:charge_grain_H2} evaluated for the parameters stated above is reported in \appx{sect:grain_H2_simple}.

\begin{figure}
   \centering
       \includegraphics[width=.49\textwidth]{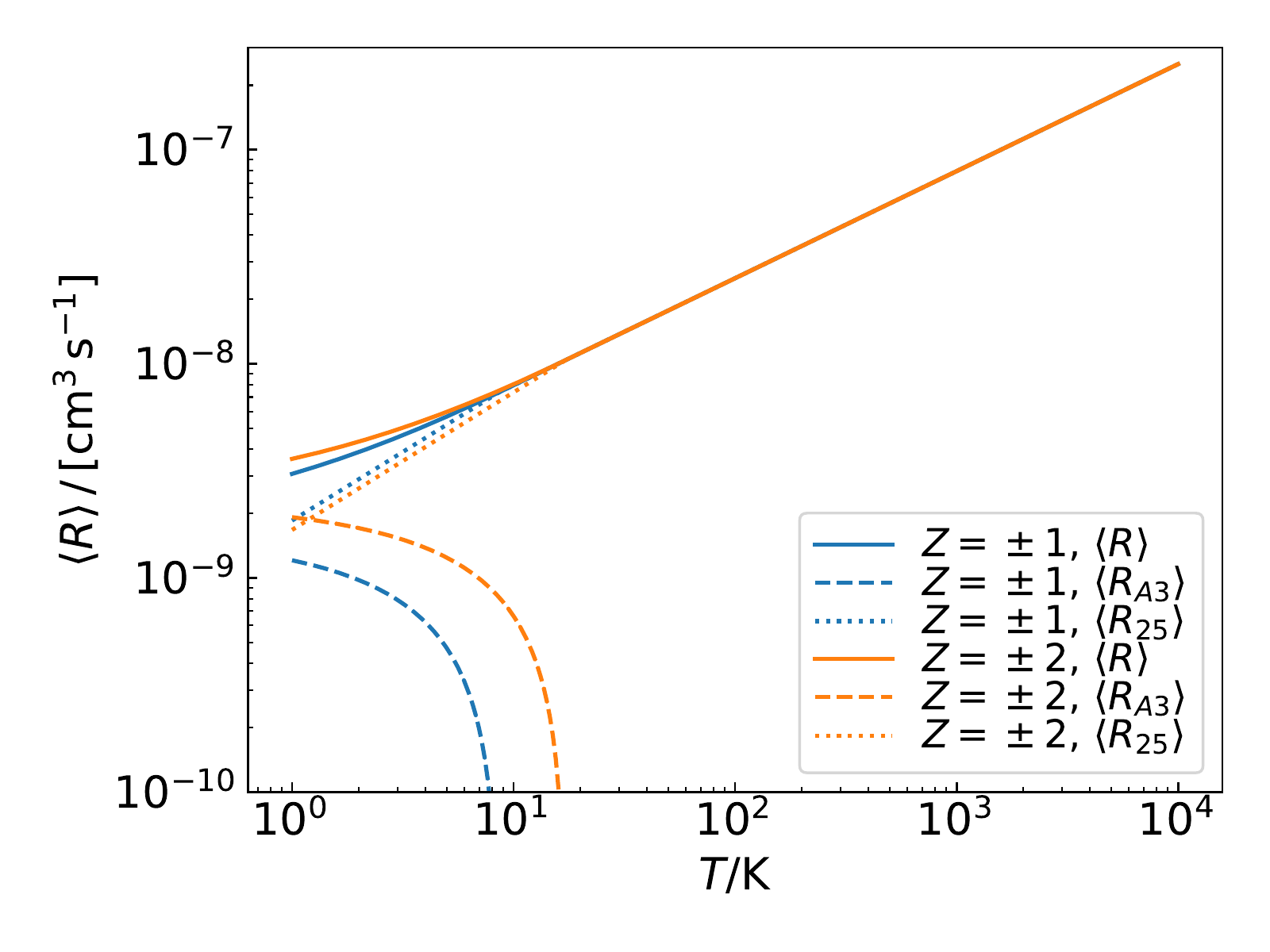}
 \caption{Temperature dependence of \eqn{eqn:charge_grain_H2}, i.e., the momentum transfer rate coefficient between charged grains and H$_2$, for different grain charge $Z$. We also plot the first ($\langle R_{\rm L}\rangle$) and the second ($\langle R_{\rm hs}\rangle$) terms of the integral. Note that $\langle R_{\rm L}\rangle$ becomes negative when $T\ge T_{\rm c}$ and hence $\langle R\rangle = \langle R_{\rm hs}\rangle$ as discussed in the text. In this Figure, we assume a power-law distribution in grain size, $\varphi(a)\propto a^p$, with $p=-3.5$ and size range $a_{\rm min}=10^{-7}$ to $a_{\rm max}=10^{-5}$~cm.}
 \label{fig:momentum_rates}
\end{figure}

We note that \eqn{eqn:charge_grain_H2} is valid only when $a_{\rm min} \le a_{\rm c}(T) \le a_{\rm max}$. Substituting $Z=1$, $\delta=1.3$, and $\alpha_{\rm pol}=8.06\times10^{-25}$~cm$^3$ into \eqn{eqn:alpha_crit}, we find that $a_{\rm c}(T=1\,{\rm K})=1.71\times10^{-7}$~cm and that it decreases as $a_{\rm c}\propto T^{-2.5}$, hence the validity of \eqn{eqn:charge_grain_H2} is only critical for $a_{\rm min}$ because $a_{\rm max}\gg a_{\rm c}$ when $T\ge 1$~K. Conversely, using the same relation, $T_{\rm c}=8.6$~K is the critical temperature corresponding to $a_{\rm min}=10^{-7}$~cm. Therefore, when $T\ge T_{\rm c}$, \eqn{eqn:charge_grain_H2} reduces to the second term on the right-hand side only (i.e.\ the normalised integral over $R_{\rm hs}$). We report the total rate and individual terms in \fig{fig:momentum_rates}. Since $Z$ does not strongly affect $\langle R_{g,n}(T,Z) \rangle$, we do not discuss the behaviour of larger $Z$ here, although we do include $Z=\pm2$ in the Figure for comparison.

\subsubsection{Electrons--H$_2$}
At low energies ($\lesssim1$~eV), the collisional rate between electrons and molecular hydrogen deviates significantly from the Langevin approximation. Therefore, we employ the fit from \citet{Pinto2008b} based on the cross-section obtained by comparing theoretical and experimental data:
\begin{eqnarray}
 R_{e,n}(T) & = & 10^{-9} \sqrt{T} \left[ 0.535 + 0.203 \log(T) - 0.163 \log(T)^2\right.\nonumber\\
       & + & \left. 0.05 \log(T)^3 \right]\,{\rm cm^3\,s^{-1}}\,,
\end{eqnarray}
where all variables are in cgs units. More details can be found in \citet{Pinto2008b}.

\subsubsection{Cations--H$_2$}
Analogously, the rate for collisions between positive ions and molecular hydrogen is also taken from \citet[A.3]{Pinto2008b}
\begin{equation}
 R_{+,n} = 2.210\,\pi\,q^2 \sqrt{\frac{\alpha_{\rm H_2}}{\mu}}\,,
\end{equation}
and is the same as discussed in \sect{sect:charged_grain_H2} when $a<a_{\rm c}$.

\section{Code structure}\label{sect:code}
In this Section, we provide a brief overview of the publicly available\footnote{\linktocode} code, \codename{}, developed for this study.

Following the approach of \krome{} \citep{Grassi2014}, the code consists of a \python{} pre-processor that computes the chemical reaction rates including dust grains, plus the momentum exchange cross-sections for the resistivity coefficient, and then writes optimised \fortran{} code that contains the MHD solver and other physics modules.

In contrast to \krome{}, in \codename{}, the \fortran{} files are directly modified by \python{} via specific directives that are recognised by the pre-processor as writable code blocks. The first stage is controlled by \file{main.py}, which creates an instance of the chemical network class (\file{network.py}) from an external file containing the reaction rate coefficients, and parses them into a set of objects according to the reaction class (\file{reaction.py}) and the species class (\file{species.py}). The reaction class also integrates any reaction rate coefficients that depend on the grain size distribution. Common variables, such as the grain size range $a_{min}$ to $a_{max}$, the exponent of the power law $\varphi(a)\propto a^{p}$, and the bulk density $\rho_0$, are defined in \file{common.py}. Dynamically-generated rate coefficient functions are written to \file{rates.f90}. The momentum exchange coefficients are also pre-computed by the pre-processor and supplied to the \fortran{} code (in \file{nonideal.f90}) via linear fitting to a logarithmically-spaced table in temperature. Finally, the \python{} pre-processor automatically generates and places the right-hand side of the chemical differential equations in \file{odechem.f90}.

The core of the second stage is the MHD solver (\file{ode.f90}), which is called by \file{test.f90}. The \file{ode.f90} file contains the call to \dlsodes\ which solves the complete set of differential equations, i.e., \eqnrange{eqn:hydro_full_first}{eqn:hydro_full_last}.

The \file{ode.f90} file also supplies the chemical differential equations (\file{odechem.f90}), the chemical reaction fluxes (\file{fluxes.f90}), and the rate coefficient constants (\file{rates.f90}) to \dlsodes. Moreover, \file{ode.f90} also accesses \file{nonideal.f90}, where the non-ideal coefficients calculation routine is contained, \file{cooling.f90} and \file{heating.f90} for cooling and heating processes, and the cosmic-ray propagation functions in \file{crays.f90}. The initial conditions for the MHD shock are stored in \file{input.dat}, while variables that remain constant during the simulation (e.g.\ $B_x$) are stored in \file{commons.f90} to help the compiler in optimise the calculation.

A single variable \verb+n(physical_variables, cells)+ is used to represent the main data structure. It includes the values of the physical variables, $\mathbf{U}$ (Eq.\ \ref{eqn:FUSsys}), for all cells. This approach allows \dlsodes\ to solve \eqnrange{eqn:hydro_full_first}{eqn:hydro_full_last} for all cells simultaneously and without any operator-splitting.

\section{Code benchmark}\label{sect:benchmark}
In this Section, we compare our resistivity calculations to \citet[][hereinafter M16]{Marchand2016}, which explores the behaviour of the non-ideal MHD coefficients using a zero-dimensional barotropic collapse problem and equilibrium chemistry. This benchmark was chosen because the source code is publicly available\footnote{\url{https://bitbucket.org/pmarchan/chemistry/}} and the setup applies to the physical regimes we are studying here, making it an ideal target for comparison.

We have, meanwhile, also successfully tested the MHD implementation in \codename{} against two additional standard benchmarks. The results can be found in \appx{sect:tests}.

Similar to \citet[][hereinafter UN90]{Umebayashi1990}, the background model of M16 is a zero-dimensional collapsing cloud with a temperature determined using the equation of state from \citet{Machida2006}, but also defined in \theireqn{9} of M16 and using the parameters listed in their \theireqn{10}. The chemical evolution of the collapse is modelled from $n_{\rm H}=10^2-10^{25}$~cm$^{-3}$, i.e., up until the formation of the second core. They set the ionisation rate to a constant $\zeta=10^{-17}$~s$^{-1}$. In contrast to UN90, the dust in their model is given a power-law size distribution $\varphi(a)\propto a^{-3.5}$. The distribution is normalised in order to obtain the same total surface area as the fiducial distribution of \citet{Kunz2009}, as shown in their Eqs.\ (16) and (17). We use the results reported in Figs.\ 3 and 5 of M16 as our reference for comparison.

The test consists of running the chemistry forward in time until equilibrium is reached at each density and then calculating the ambipolar diffusion ($\eta_{\rm AD}$), Ohmic resistivity ($\eta_{\rm O}$), and Hall resistivity ($\eta_{\rm H}$) coefficients. We initialise the temperature and magnetic field as functions of number density following M16. In our case, we limit the density range to $1-10^{12}$~cm$^{-3}$ because we miss some physical processes that are important at higher densities/temperatures, e.g., grain sublimation (see Fig.\ 2 of M16). Grains are permitted have a charge from $Z=-2$ to $Z=2$.

In order to reproduce their results, we use a grain size distribution $\varphi\propto a^{-3.5}$ with size range $a_{\rm min}=1.81\times10^{-6}$ to $a_{\rm max}=9.049\times10^{-5}$~cm, and a dust-to-gas ratio of $\mathcal{D}=0.1$. We also adopt the recombination rate coefficient from M16
\begin{equation}\label{eqn:recombination_M16}
 k_{\rm rec}(T)=2.4\times10^{-7}\left( T/300\,{\rm K} \right)^{-0.69}\,{\rm cm^{3}\,s^{-1}}
\end{equation}
instead of our \eqn{eqn:gas_recombination}. In this test we assume sticking coefficient $S=1$ for all the rates involving grains, except for the electron-neutral grain attachment where $S=0.6$. Moreover, we do not use the Mg$^+$ reduced mass, but the actual one.

\begin{figure}
   \centering
       \includegraphics[width=.49\textwidth]{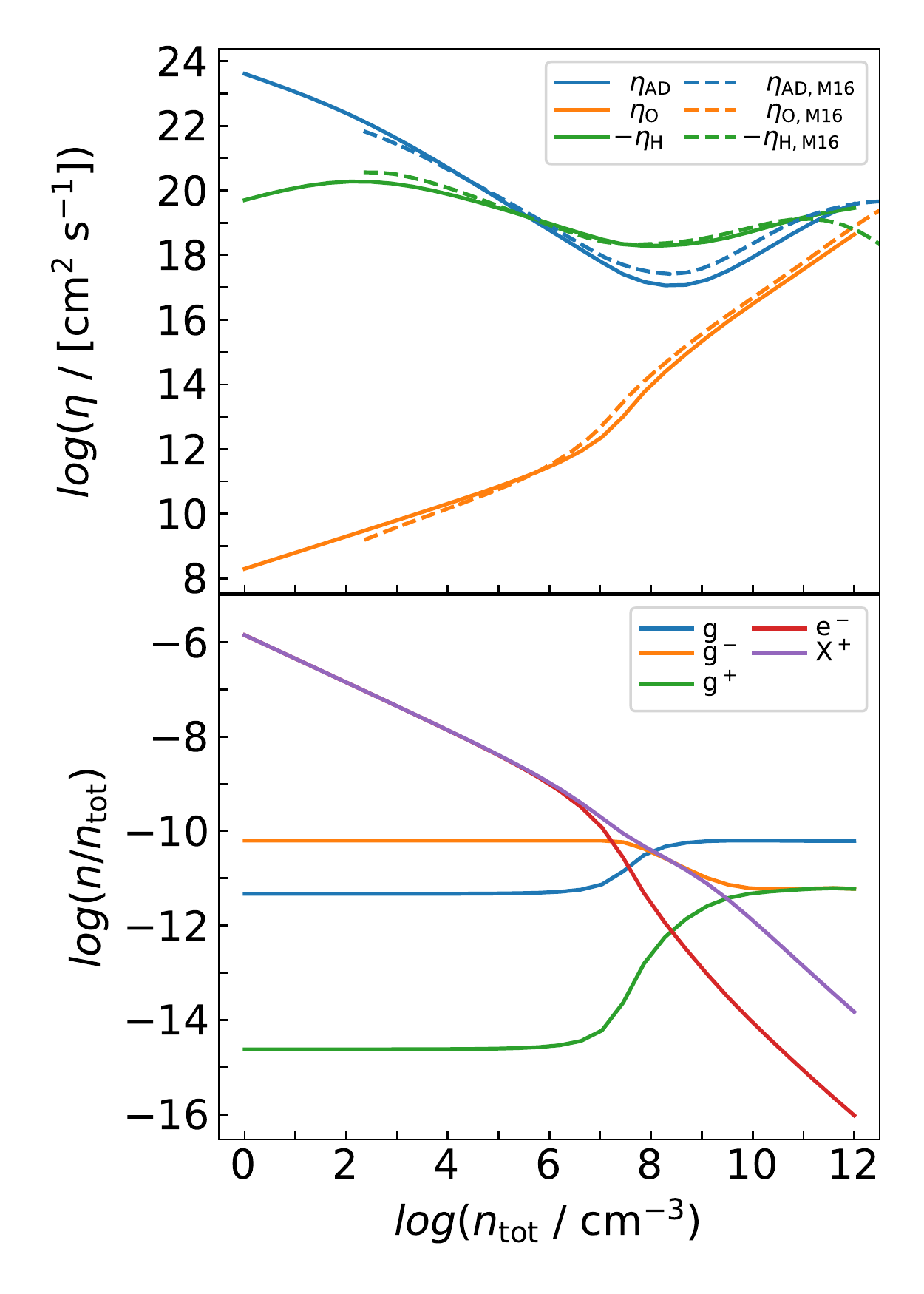}
 \caption{Top: Resistivity coefficient results for the barotropic collapse test of \citet{Marchand2016} (dashed) and our code (solid). This Figure should be compared with their Fig.\ 5. Bottom: Corresponding fractional abundances of neutral (g) and charged grains (g$^{\pm}$), electrons (e$^-$), and cations X$^+$. For the sake of clarity, we omit g$^{--}$ and g$^{++}$ from the plot. The behaviour is identical to that found in the Appendix of \citet{Marchand2016}. In both panels, $n_{\rm tot}$ is the initial total number density.}
 \label{fig:M16}
\end{figure}

We report out results in \fig{fig:M16}, where we find good agreement with M16 in both chemical abundances and resistivity coefficients. See also \appx{appx:M16} for further details.

\section{The effects of microphysics on the structure of magnetic shocks}\label{sect:shock_parameters}
The aim of this study is to understand the effects of chemistry, dust microphysics, cosmic rays, and cooling/heating on the evolution of astrophysical magnetic shocks. In this Section, we analyse these effects in detail by varying the physical ingredients and parameters of a reference model. The different models and their characteristics are reported in \tab{tab:tests_report}.

\newcommand{\test}[1]{\texttt{#1}}

\begin{table*}
\begin{tabular}{lll}
\hline
Model name & Description & Reference parameter or model\\
\hline
\test{reference} & see \sect{sect:reference_model} & ---\\
\test{reference\_noth} & without radiative cooling or CR heating & \test{reference}\\
\test{ideal} & ideal MHD & \test{reference}\\
\test{ideal\_noth} & \test{ideal} without radiative cooling or CR heating & \test{ideal}\\
\test{adtab} & $\eta_{\rm AD}$ from equilibrium tables & time-dependent chemistry\\
\test{arec} & alternative recombination rate (Eq.\ \ref{eqn:recombination_M16})  & recombination rate from \eqn{eqn:gas_recombination}\\
\test{stick01} & electron sticking $S(T)=0.1$ & fit to \citet{Bai2011}; see \sect{sect:sticking}\\
\test{stick1} & electron sticking $S(T)=1$ & fit to \citet{Bai2011}; see \sect{sect:sticking}\\
\test{N5e20} & initial effective column density $N_{{\rm eff},1}=5\times10^{20}$~cm$^{-2}$ & $N_{{\rm eff},1}=5\times10^{21}$~cm$^{-2}$\\
\test{N5e22} & initial effective column density $N_{{\rm eff},1}=5\times10^{22}$~cm$^{-2}$ & $N_{{\rm eff},1}=5\times10^{21}$~cm$^{-2}$\\
\test{cr}$x$ & constant cosmic ray ionisation rate $\zeta=10^{-x}$ & cosmic ray attenuation as in \sect{sect:cosmic_rays}\\
\test{crlow} & ``low'' CRs; $b_1=1.327\times10^{-12}$~s$^{-1}$ and $b_2=-0.211$ & ``high'' CRs; $b_1=5.34\times10^{-5}$~s$^{-1}$ and $b_2=-0.384$\\
\test{bulk10} & dust bulk density $\rho_0=10$~g~cm$^{-3}$ & $\rho_0=3$~g~cm$^{-3}$\\
\test{d2g1e4} & dust to gas mass ratio $\mathcal{D}=10^{-4}$ & $\mathcal{D}=10^{-2}$\\
\test{d2g\_step} & $\mathcal{D}=10^{-6}$ initially in the upstream region, $\mathcal{D}=10^{-2}$ otherwise & $\mathcal{D}=10^{-2}$\\
\test{nogg} & no grain-grain chemistry & grain-grain chemistry included\\
\test{pexp25} & $p=-2.5$ in $\varphi(a)\propto a^p$ & $p=-3.5$\\
\test{pexp50} & $p=-5$ in $\varphi(a)\propto a^p$ & $p=-3.5$\\
\test{amin1e6} & $a_{\rm min}=10^{-6}$~cm in $\varphi(a)\propto a^p$ & $a_{\rm min}=10^{-7}$~cm\\
\test{cool01} & cooling rate multiplied by $0.1$ & standard cooling\\
\test{cool10} & cooling rate multiplied by $10$ & standard cooling\\
\test{noheat} & no cosmic ray heating & cosmic ray heating included\\
\test{nochem1e}$x$ & chemistry not solved; constant ionisation fraction $f_i=10^{-x}$ & time-dependent chemistry, ionisation fraction\\
\hline
\end{tabular}\caption{Description and parameters of the different models along with the corresponding default/comparison. See the text for additional details.}
\label{tab:tests_report}
\end{table*}

\subsection{Reference model}\label{sect:reference_model}
\begin{figure}
   \centering
       \includegraphics[width=.49\textwidth]{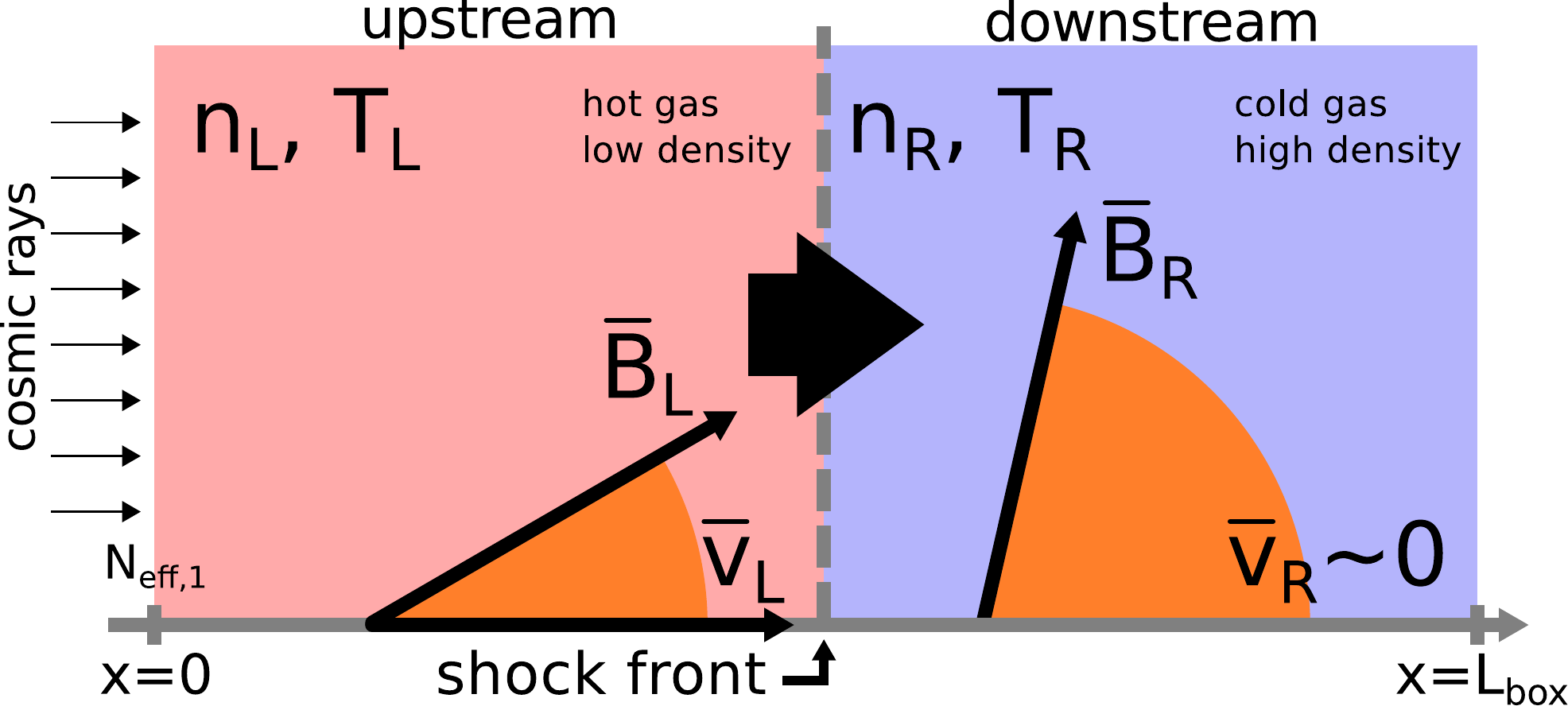}
 \caption{Sketch of the initial shock tube conditions. The shocked gas (hot, fast, low density) moves from left to right, colliding with unperturbed gas (cold, nearly stationary, high density). Cosmic rays enter the simulation box at the left edge with initial effective column density $N_{{\rm eff},1}$ and propagate following \sect{sect:cosmic_rays}. The magnetic field pitch angles are exaggerated for the sake of clarity.}
 \label{fig:shock_sketch}
\end{figure}

All of the tests presented in this Section are based on a \mbox{1-D} MHD \test{reference} shock tube model with a box size of $L_{\rm box}=3\times 10^{17}$~cm ($\simeq 0.1$~pc~$\simeq 2\times10^4$~AU) and $1024$ linearly-spaced grid points. The shock moves from left to right (see \fig{fig:shock_sketch} and \tab{tab:initial_conditions}), with initial left state density $n_{\rm L}=10^4$~cm$^{-3}$, velocity components $v_{x,{\rm L}}=10^6$~cm~s$^{-1}$ and $v_{y,{\rm L}}=0$~cm~s$^{-1}$, magnetic field $B_{y,{\rm L}}=10^{-4}$~G, and temperature $T_{\rm L}=10^3$~K. The unperturbed right side has $n_{\rm R}=4\times10^4$~cm$^{-3}$, velocity components $v_{x,{\rm R}}=v_{y,{\rm R}}=10$~cm~s$^{-1}$ (note $\mathbf{v}_{\rm L}\gg \mathbf{v}_{\rm R}$), $B_{y,{\rm R}}=2\times10^{-4}$~G, and $T_{\rm R}=10$~K. Both sides have $B_{x,{\rm L}}= B_{x,{\rm R}}=10^{-4}$~G constant in time, $v_{z,{\rm L}}=v_{z,{\rm R}}=0$, and $B_{z,{\rm L}}=B_{z,{\rm R}}=0$. The interface between left and right states is placed at $L=0.3\,L_{\rm box}$. In all tests, we let the system evolve for $t=10^4$~yr.

\begin{table}
\begin{tabular}{llll}
\hline
Variable & Left state & Right state & Units\\
\hline
 $n$ & $10^4$ & $4\times10^4$ & cm$^{-3}$\\
 $T$ & $10^3$ & $10$ & K\\
 $B_x$ & $10^{-4}$ & $10^{-4}$ & G\\
 $B_y$ & $10^{-4}$ & $2\times10^{-4}$ & G\\
 $B_z$ & $0$ & $0$ & G\\
 $v_x$ & $10^6$ & $10$ & cm~s$^{-1}$\\
 $v_y$ & $0$ & $10$ & cm~s$^{-1}$\\
 $v_z$ & $0$ & $0$ & cm~s$^{-1}$\\
 $\mathcal{D}$ & $10^{-2}$ & $10^{-2}$ & ---\\
 $f_i$ & $10^{-7}$ & $10^{-7}$ & ---\\
\hline
\end{tabular}\caption{Initial conditions for the MHD shock tube model.}
\label{tab:initial_conditions}
\end{table}

The \test{reference} model includes the full calculation of ambipolar diffusion, time-dependent chemistry with the recombination rate coefficient from \eqn{eqn:gas_recombination}, and electron-sticking following \citet{Bai2011}. We set the dust-to-gas mass ratio to $\mathcal{D}=\rho_d/\rho=10^{-2}$ and the grain size distribution to $\varphi(a)\propto a^p$ with $p=-3.5$ and size range $a_{\rm min}=10^{-7}$~cm to $a_{\rm max}=10^{-5}$~cm. The dust is given a bulk density of $\rho_0=3$~g~cm$^{-3}$. The initial ionisation fraction is set to $f_i = n_{\rm e^-} / n_{\rm H_2} = 10^{-7}$, but this has no influence on the evolution except for tests without chemistry (i.e.\ \test{nochem1e}$x$). See \appx{sect:initial} for additional details on the chemical initial conditions.

Since the gas is dominated by molecular hydrogen, we assume a molecular gas with constant $\gamma=7/5$ and constant mean molecular weight $\mu=2$. Cosmic rays propagate from left to right with initial effective column density $N_{{\rm eff},1}=5\times10^{21}$~cm$^{-2}$, and we assume \emph{high} cosmic ray spectrum, i.e., \eqn{eqn:cr_fit} with $b_1=5.34\times10^{-5}$~s$^{-1}$ and $b_2=-0.384$ (\sect{sect:cosmic_rays}).

\subsection{General behaviour of the models}\label{sect:overall_behaviour}

\begin{figure*}
   \centering
       \includegraphics[width=.99\textwidth]{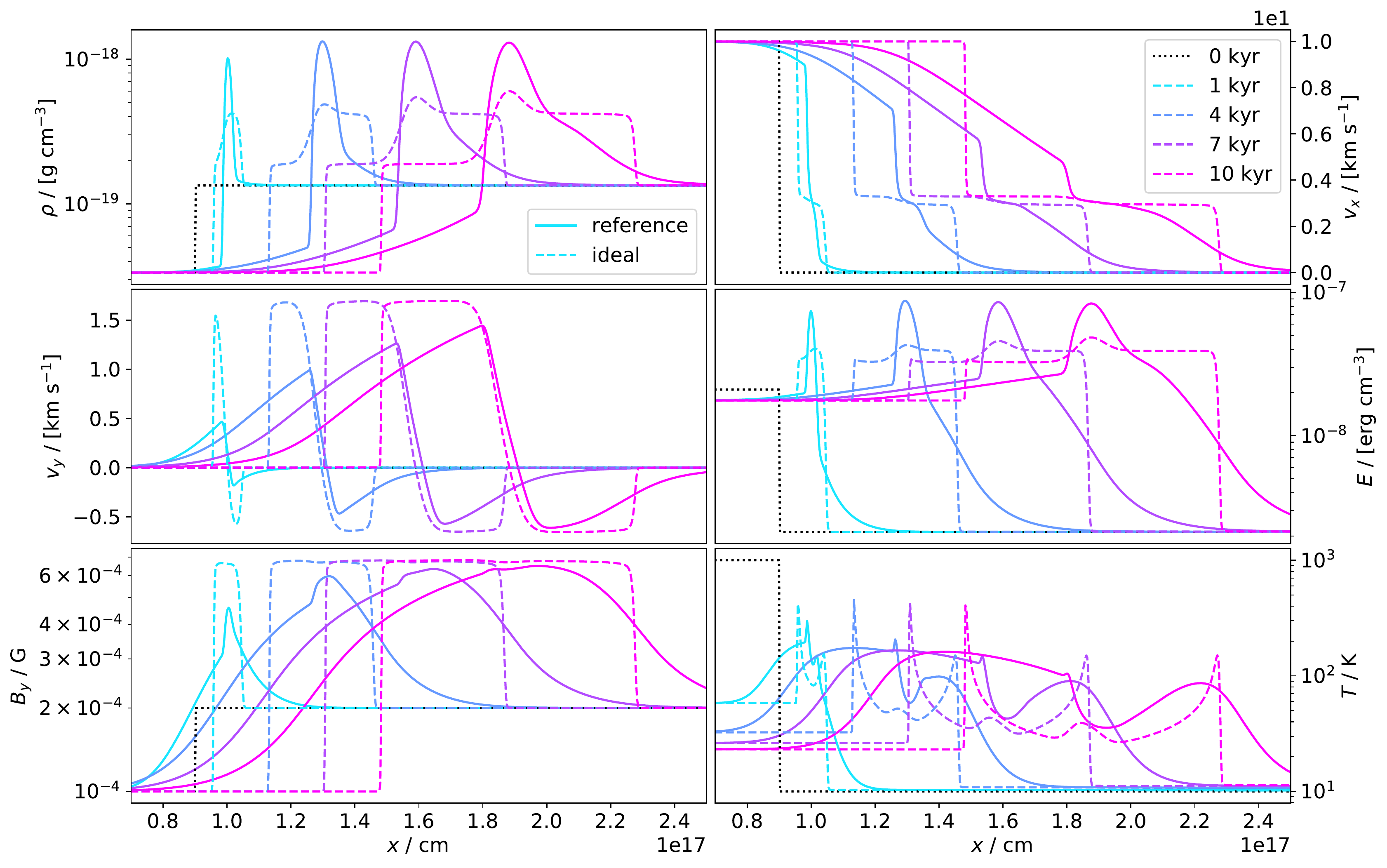}
 \caption{Temporal evolution of the MHD shock. Plotted are density ($\rho$), $x$- and $y$-components of the velocity ($v_x$, $v_y$), $y$-component of the magnetic field ($B_y$), energy density ($E$), and temperature ($T$) for the \test{reference} and \test{ideal} models. The curves are plotted at $t=1, 4, 7$ and $10$~kyr from left to right. The black dotted lines indicate the initial conditions (which are the same for both models).}
 \label{fig:time_evolution}
\end{figure*}

\begin{figure*}
   \centering
       \includegraphics[width=.99\textwidth]{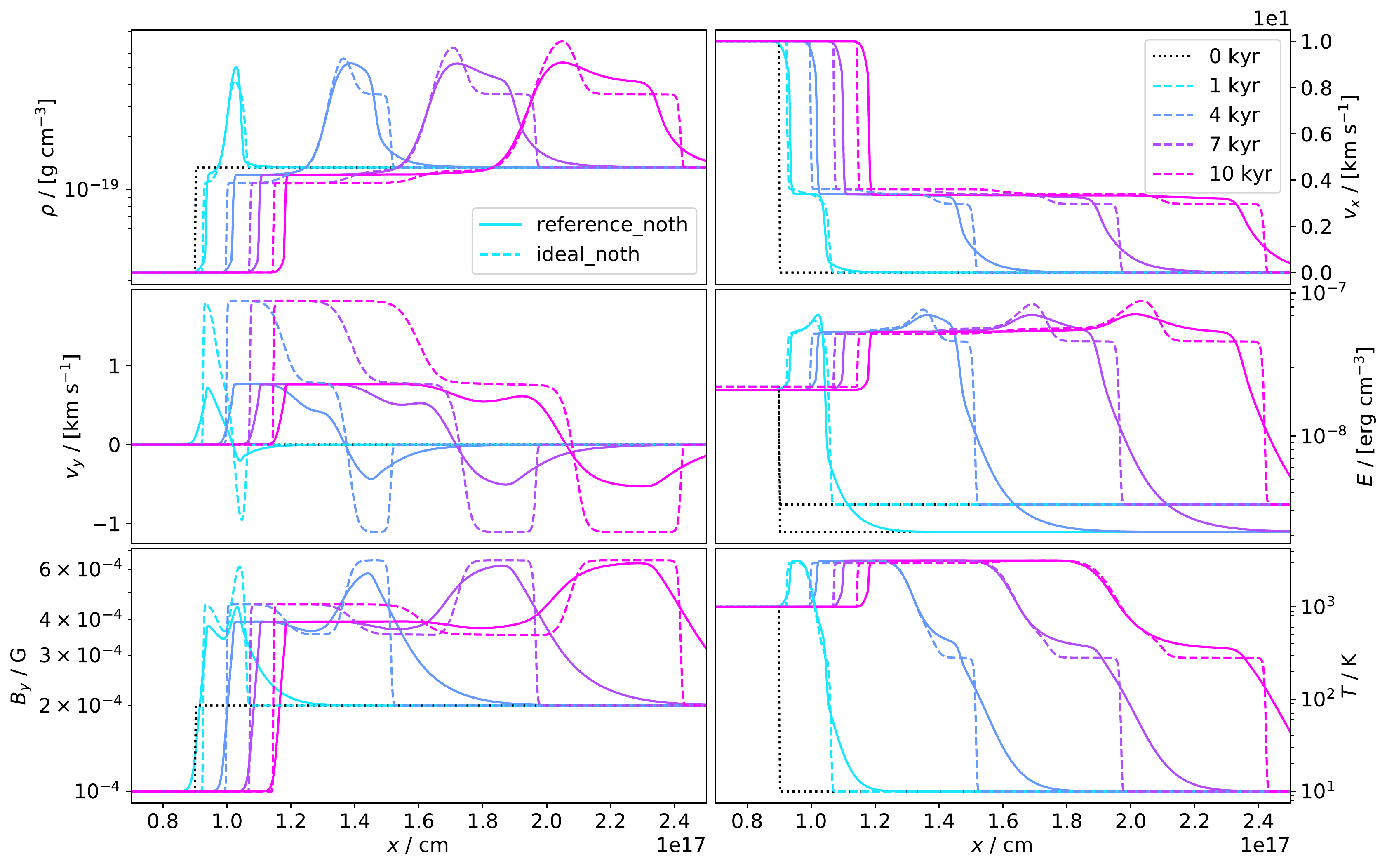}
 \caption{Temporal evolution of the MHD shock \emph{without} cooling and heating ($\Lambda=\Gamma=0$). Plotted are density ($\rho$), $x$- and $y$-components of the velocity ($v_x$, $v_y$), $y$-component of the magnetic field ($B_y$), energy density ($E$), and temperature ($T$) for the \test{reference\_noth} and \test{ideal\_noth} models. The curves are plotted at $t=1, 4, 7$ and $10$~kyr from left to right. The black dotted lines indicate the initial conditions (which are the same for both models).}
 \label{fig:time_evolution_nothermo}
\end{figure*}

We evolve the shock for $t=10^4$~yr and, in Figs.\ \ref{fig:time_evolution} and \ref{fig:time_evolution_nothermo}, we report\footnote{An animation of the evolution of \test{reference} and \test{ideal} models is available at \url{https://vimeo.com/286491689}; the evolution without thermal processes (\test{reference\_noth}, \test{ideal\_noth}) is available at \url{https://vimeo.com/290131160}.} the results for \test{reference}, \test{ideal}, \test{reference\_noth} and \test{ideal\_noth} models.

The solution to the \test{ideal\_noth} test, which is the simplest physical model we explore, exhibits five features (from left to right): a fast shock, a slow rarefaction, a contact discontinuity, a slow shock, followed by another fast shock. While the leftmost and rightmost fast shocks are clear, the slow rarefaction, located near $\sim1.8\times 10^{17}$~cm at 10~kyr, is very low amplitude and only just visible in $v_y$. The contact discontinuity and slow shock are, meanwhile, adjacent to each other with the transition between the two occurring at $\sim2\times 10^{17}$~cm (at 10~kyr). The solution as a whole moves to the right at $\sim3.5$~km~s$^{-1}$. The $z$-components of the magnetic field and velocity ($B_z$ and $v_z$) remain equal to zero since both are initially zero and there are no $z$-derivatives in the problem. Thus, the magnetic field vector does not rotate for these particular initial conditions, even though the code is capable of this.

The \test{ideal} model, which includes cooling (\sect{sect:cooling}), demonstrates a considerably different solution. The shock structure is modified significantly by the cooling of the hot gas in the left initial state ($T_{\rm L} = 10^3$~K $\rightarrow$ $\sim 25$~K) and the shock-heated gas in the intermediate region between left and right fast shocks (cf.\ the \test{ideal\_noth} model). The wave speeds and the extent of the region between the leftmost fast shock and the contact discontinuity/slow shock are subsequently reduced. Even the self-similarity of the \test{ideal\_noth} solution is broken.

The addition of ambipolar diffusion smears out gradients in the magnetic field ($B_y$ in this case) and subsequently heats the affected regions. This is particularly visible for the region downstream from the rightmost fast shock in the \test{reference\_noth} model (\fig{fig:time_evolution_nothermo}). Note, however, that the diffusion of the magnetic field in the vicinity of the leftmost fast shock only becomes substantial once cooling is included (i.e.\ the \test{reference} model); this is a result of recombination, which is more efficient at lower temperatures (\fig{fig:grain_rates}), and reduces the ionisation fraction. That said, even including ambipolar heating, magnetic diffusion does not drastically modify the structure and evolution of the solution; cooling has a much more significant effect.

Before turning to a more detailed examination of how the microphysics affects the shock solution, we first describe our primary means of presenting and comparing the different tests. Figure \ref{fig:coloramagram_rho} reports the density in the different models at $t=10^{4}$~yr for the region $x=10^{17}$ -- $2.4\times10^{17}$~cm. The absolute value of the gas density is plotted in the four top panels, where the black solid line indicates the \test{reference} model, while the shaded grey area denotes the envelope values/extrema of the density across all models for visual reference. The lower panels show the relative density variation $r_{\rho}=(\rho_{\rm model}-\rho_{\rm reference})/\rho_{\rm reference}$ for each model, sorted from the largest $|r_{\rho}|$ to the smallest.

The upper bound of the grey envelope in \fig{fig:coloramagram_rho} is mainly set by the \test{ideal} MHD test, which has the largest positive $r_{\rho}$. Conversely, the \test{nochem1e7} model, with constant ionisation fraction $f_i=10^{-7}$, gives the largest negative $r_{\rho}$ values and is mostly responsible for the lower bound of the envelope. In the region of the slow shock ($x \sim 1.9\times 10^{17}$~cm), however, it is instead the \test{ideal} model which sets the lower bound of the envelope and the \test{nochem1e7} model which sets the upper bound.

Analogously, Figs.\ \ref{fig:coloramagram_B}, \ref{fig:coloramagram_T}, and \ref{fig:coloramagram_eAD} report, respectively, on the modulus of the magnetic field ($\mathbf{B}$), the temperature ($T$), and the resistivity coefficient ($\eta_{AD}$) as well as the relative differences with respect to the \test{reference} model.

\newcommand{\captionrama}[2]{Comparison of the #1 profiles at the end of the simulation ($t=10^4$~yr) between the different models described in \tab{tab:tests_report}. The first four panels report $#2(x)$ in the different models, \emph{reference} indicates the complete model, and the shaded grey area is the envelope/extrema for all the models. The lower panels report the relative difference. The labels indicate the corresponding model, ``min'' and ``max'' the minimum and maximum values of $r_#2$, and the vertical dashed line where $r_#2=0$. A legend connecting the lower panels to the different lines in the upper four panels is provided. Panels are sorted by descending maximum value of $|r_#2|$.}

\begin{figure*}
   \centering
       \includegraphics[width=.98\textwidth]{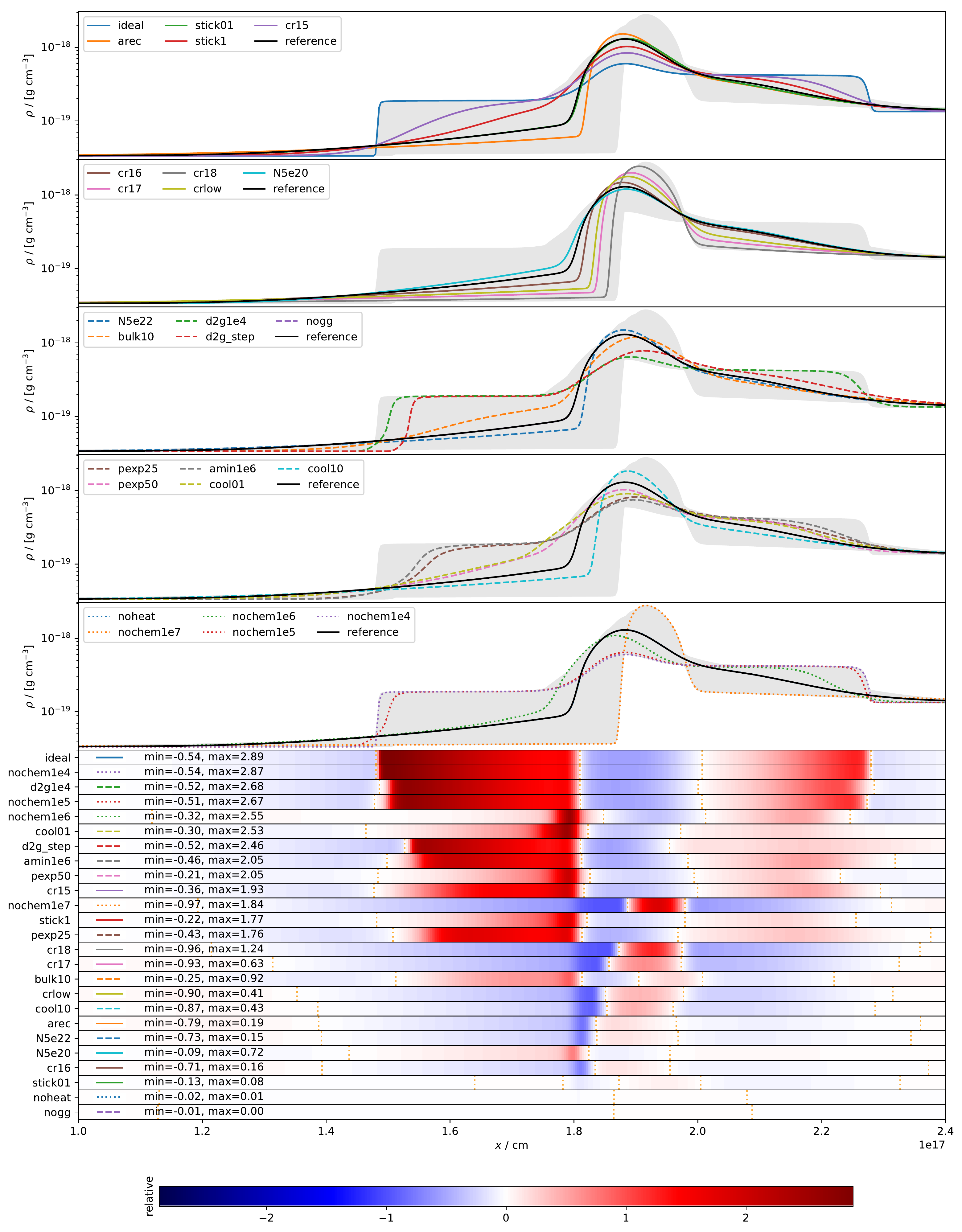}
 \caption{\captionrama{density}{\rho}}
 \label{fig:coloramagram_rho}
\end{figure*}

\begin{figure*}
   \centering
       \includegraphics[width=.98\textwidth]{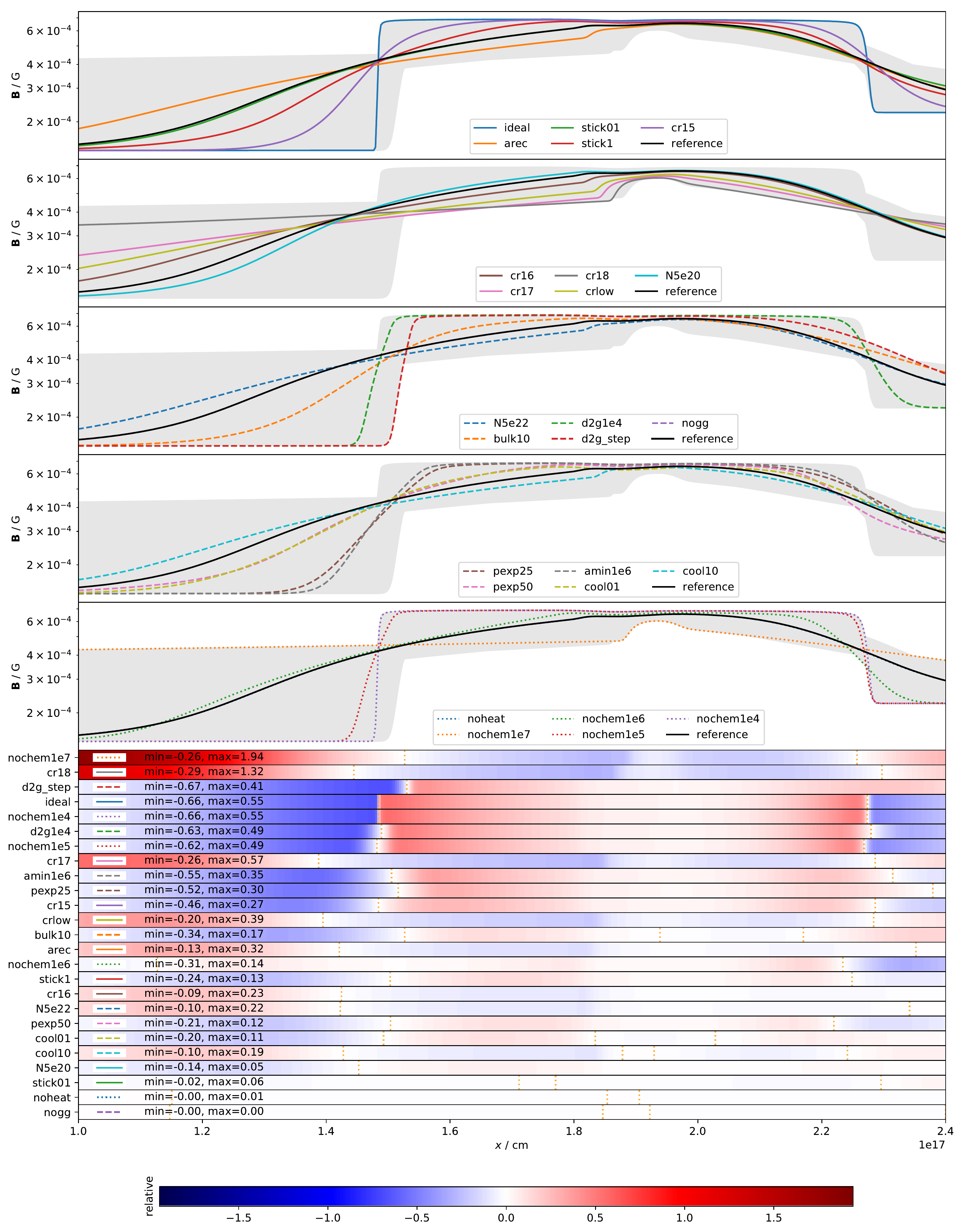}
 \caption{\captionrama{magnetic field}{\mathbf{B}}}
 \label{fig:coloramagram_B}
\end{figure*}

\begin{figure*}
   \centering
       \includegraphics[width=.98\textwidth]{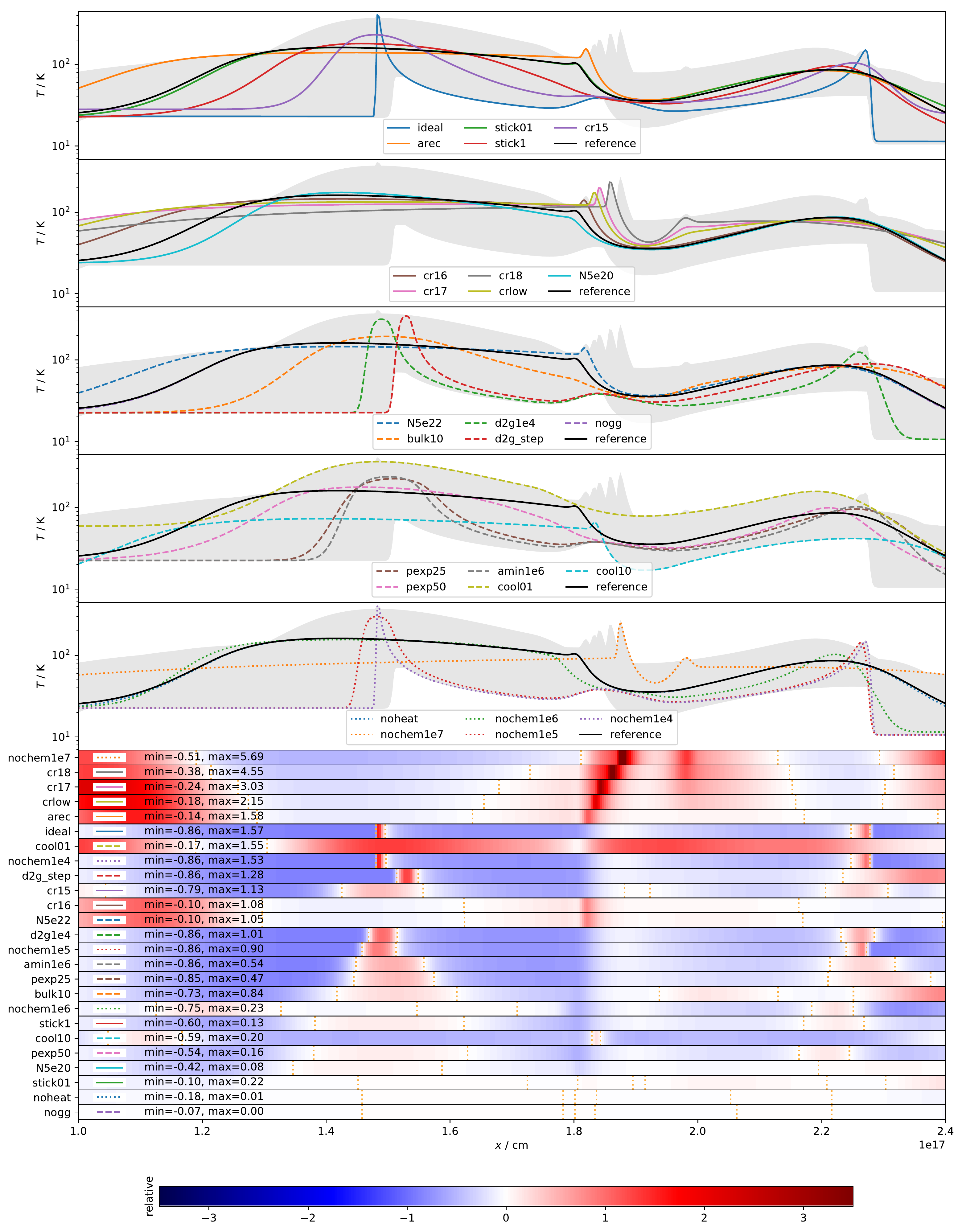}
 \caption{\captionrama{temperature}{T}}
 \label{fig:coloramagram_T}
\end{figure*}

\begin{figure*}
   \centering
       \includegraphics[width=.98\textwidth]{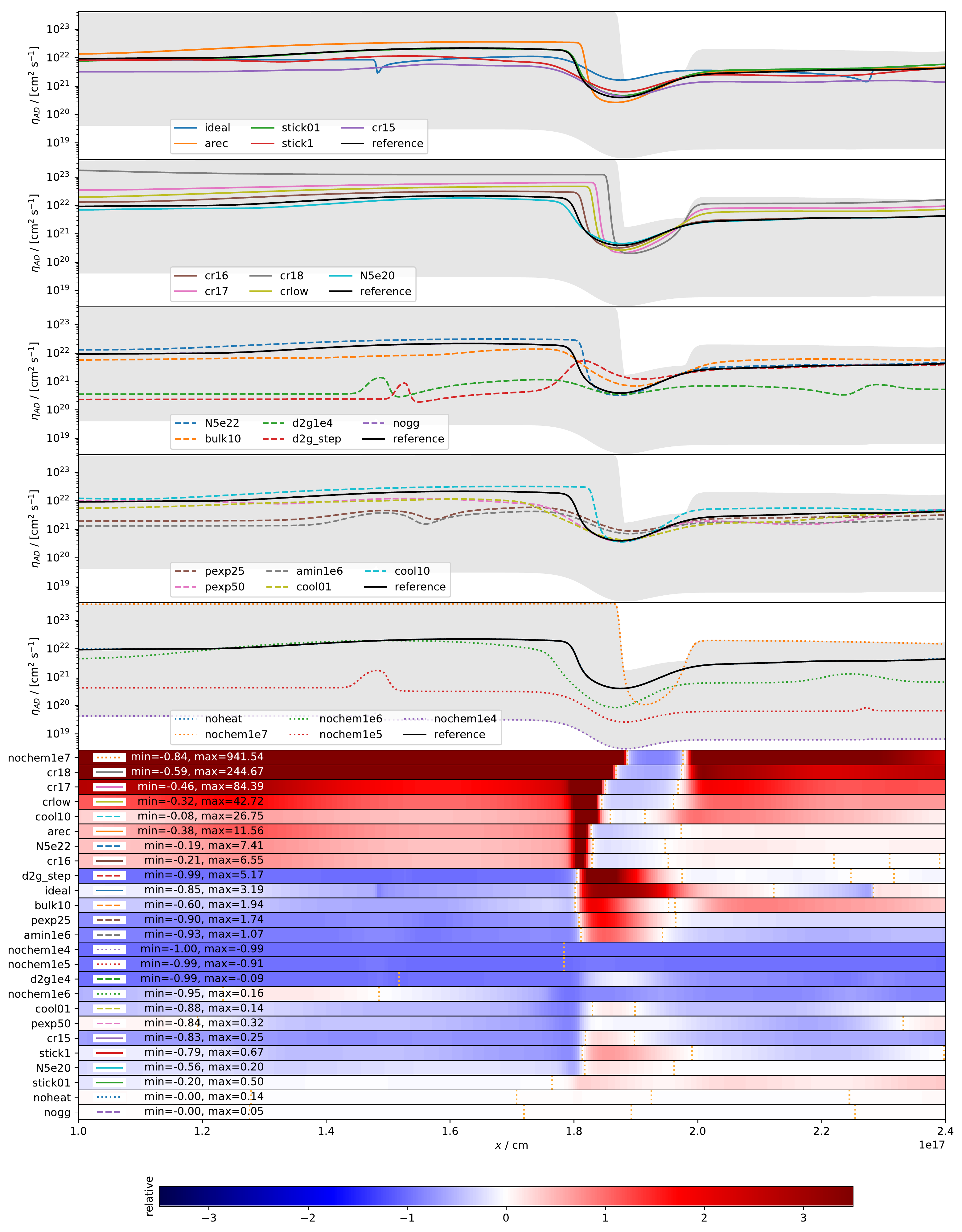}
 \caption{\captionrama{resistivity coefficient}{{\eta_{\rm AD}}}}
 \label{fig:coloramagram_eAD}
\end{figure*}

\subsection{The effect of modified chemistry}\label{sect:vary_chemistry}
Chemistry plays a key role in the evolution of the shock because the abundances of the ions that control the Hall parameter $\beta_{i,n}$, and hence the resistivity coefficient $\eta_{\rm AD}$ (see \sect{sect:beta}), are determined by the chemistry. Indeed, $\beta_{i,n}$ and $\eta_{\rm AD}$ provide the main interplay between chemistry and hydrodynamics (see \fig{fig:scheme_all}). To understand which chemical processes are most influential to the shock evolution, we change the recombination efficiency (model \test{arec}), the electron sticking (models \test{stick01} and \test{stick1}), or turn off the chemistry (\test{nochem1e4}, \test{nochem1e5}, \test{nochem1e6}, and \test{nochem1e7}). As already discussed, parameters that reduce the ionisation fraction should shift models towards stronger ambipolar diffusion.

\subsubsection{Cation-electron recombination rate}
In our model, cations only recombine with electrons via $k_{\rm rec}$ (Eq.\ \ref{eqn:gas_recombination}). Positively-charged grains and electrons can ``recombine'', but this is an aspect of the grain chemistry and sticking coefficient, which is discussed below. When we adopt an alternative $k_{\rm rec}$ (Eq.\ \ref{eqn:recombination_M16}; the \test{arec} model), we note that the solution becomes ``less ideal'' relative to the \test{reference} model (e.g.\ \fig{fig:coloramagram_rho}). The recombination rate in the \test{arec} model is much more effective than the one used in the \test{reference} model. Thus, the \test{arec} model has a lower abundance of free electrons, which diminishes the global ionisation fraction and enhances the magnetic diffusion (\fig{fig:coloramagram_eAD}).

Given our reduced network, wherein cations are represented by Mg$^+$, \eqn{eqn:gas_recombination} is the most appropriate recombination rate to use. Although the \test{arec} model demonstrates that one must be careful when choosing an effective recombination rate for reduced networks, Figs.\ \ref{fig:coloramagram_rho} -- \ref{fig:coloramagram_eAD} instead show that it is not the most important effect in determining the evolution and structure of the shock.

\subsubsection{Electron-grain sticking coefficient}\label{sect:vary_sticking}
The electron-grain sticking coefficient, $S(T)$, dictates the likelihood that electrons attach to grains after a collision. This not only affects the abundance of free electrons, but also the fraction of negatively-charged grains. Since the grain-cation sticking coefficient is typically greater than the electron-cation rate (see \fig{fig:grain_rates}), increasing the fraction of negatively charged grains then increases the probability that Mg$^+$ recombines with grains.

\begin{figure}
   \centering
       \includegraphics[width=.48\textwidth]{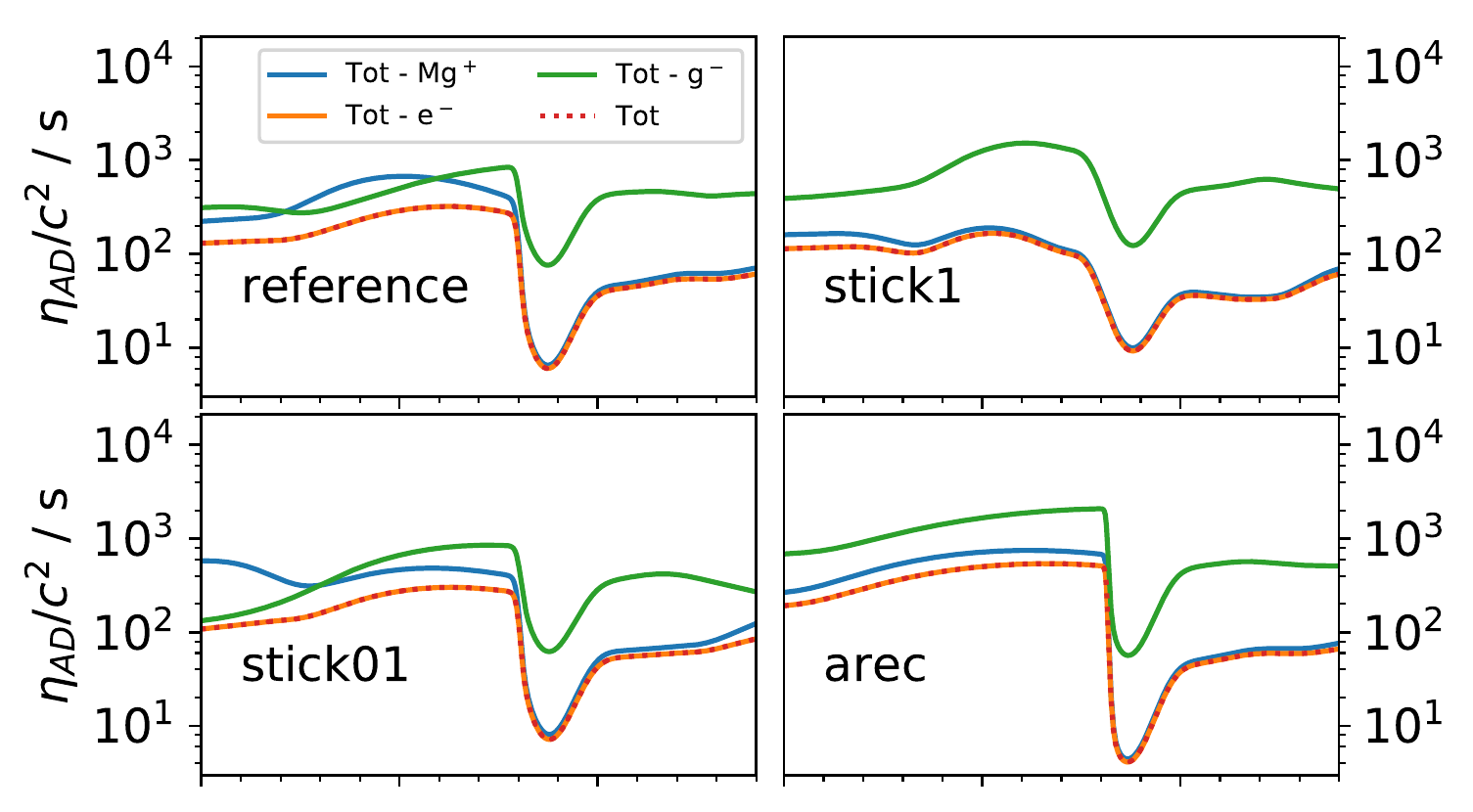}

        \vspace{-.35cm}

       \includegraphics[width=.48\textwidth]{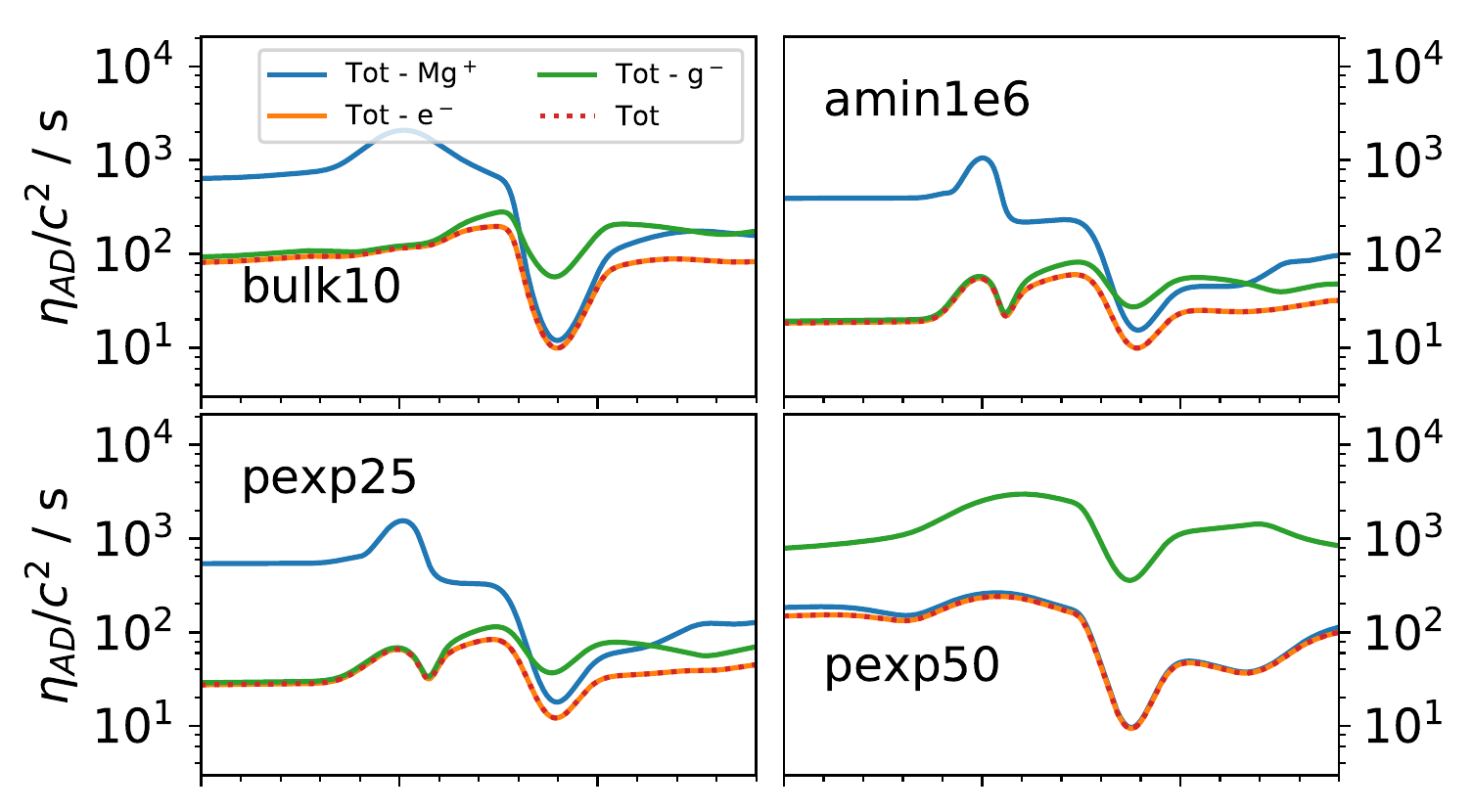}

        \vspace{-.35cm}

       \includegraphics[width=.48\textwidth]{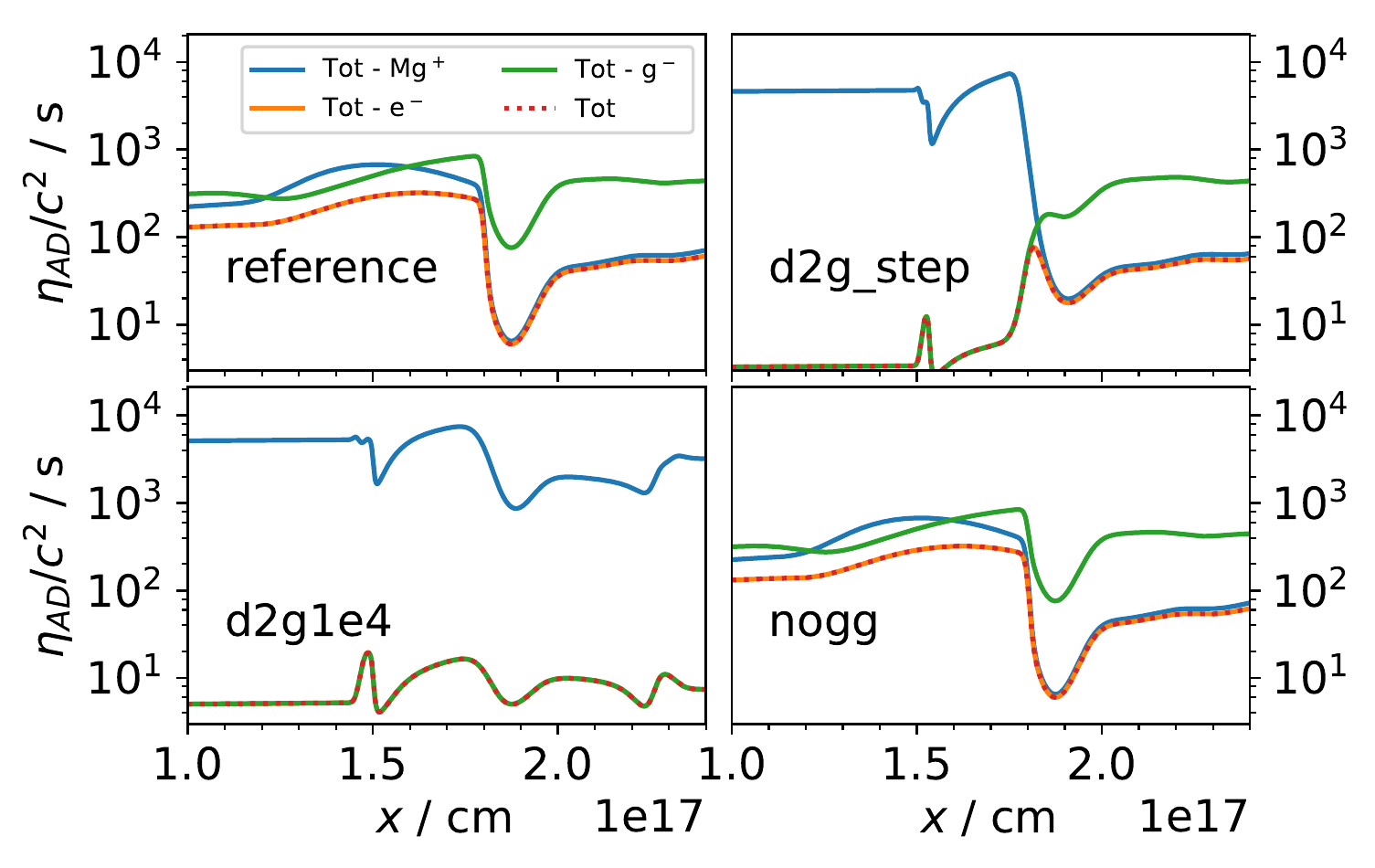}
 \caption{Comparison of the contributions to $\eta_{\rm AD}$ by different ions in different models (indicated in each panel). Each curve is the \emph{Tot}al $\eta_{\rm AD}$ computed \emph{without} the ion indicated in the legend (e.g.\ \emph{Tot - Mg}$^+$ is $\eta_{\rm AD}$ computed assuming $\rho_{\rm Mg^+}=0$). The larger the distance of a curve from \emph{Tot} (dashed red line), the more significant the contribution of the corresponding charged species. Contributions from other charged species are omitted because they are not significant.}
 \label{fig:max_eAD}
\end{figure}

\begin{figure}
   \centering
       \includegraphics[width=.48\textwidth]{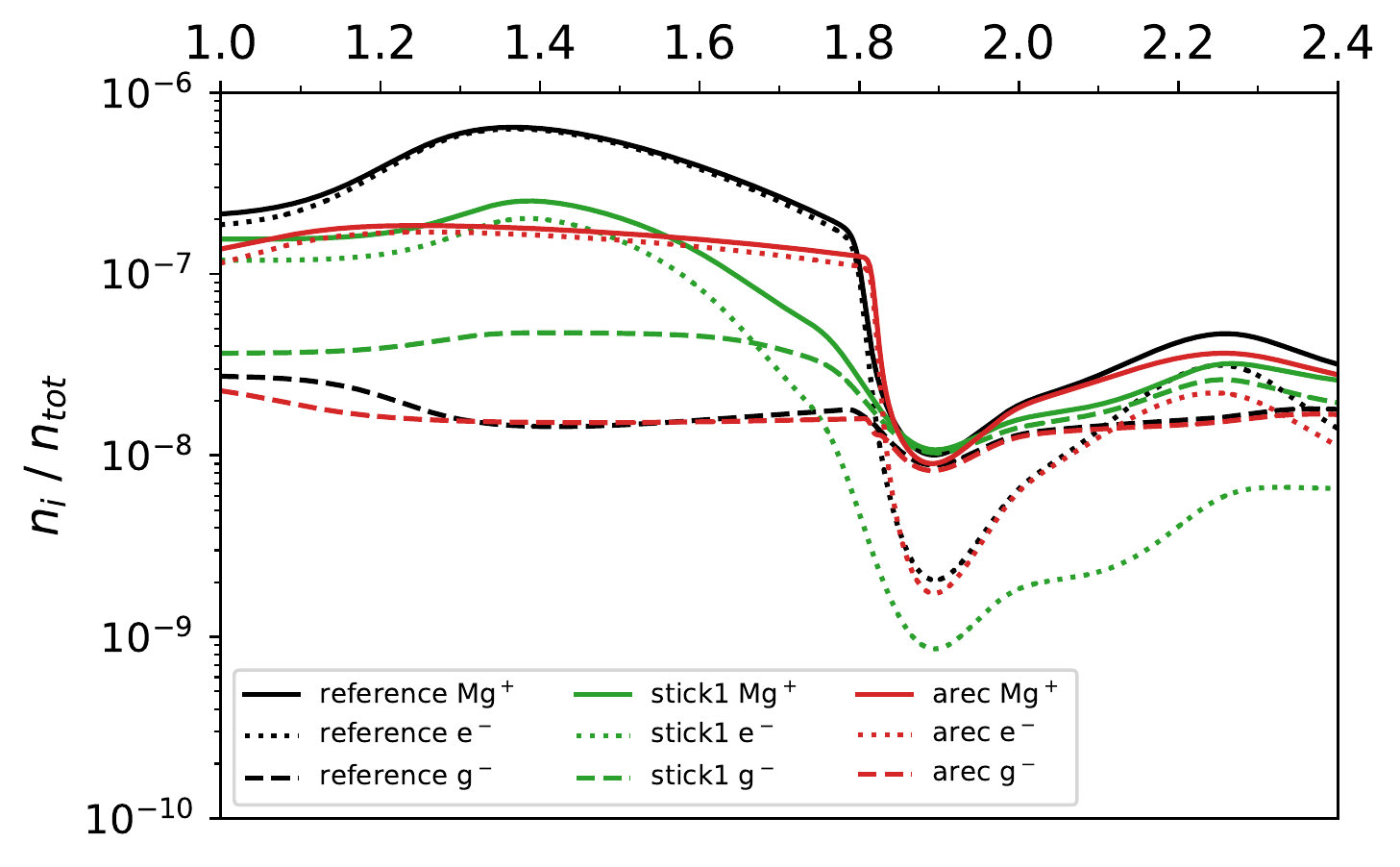}

        \vspace{-.4cm}

       \includegraphics[width=.48\textwidth]{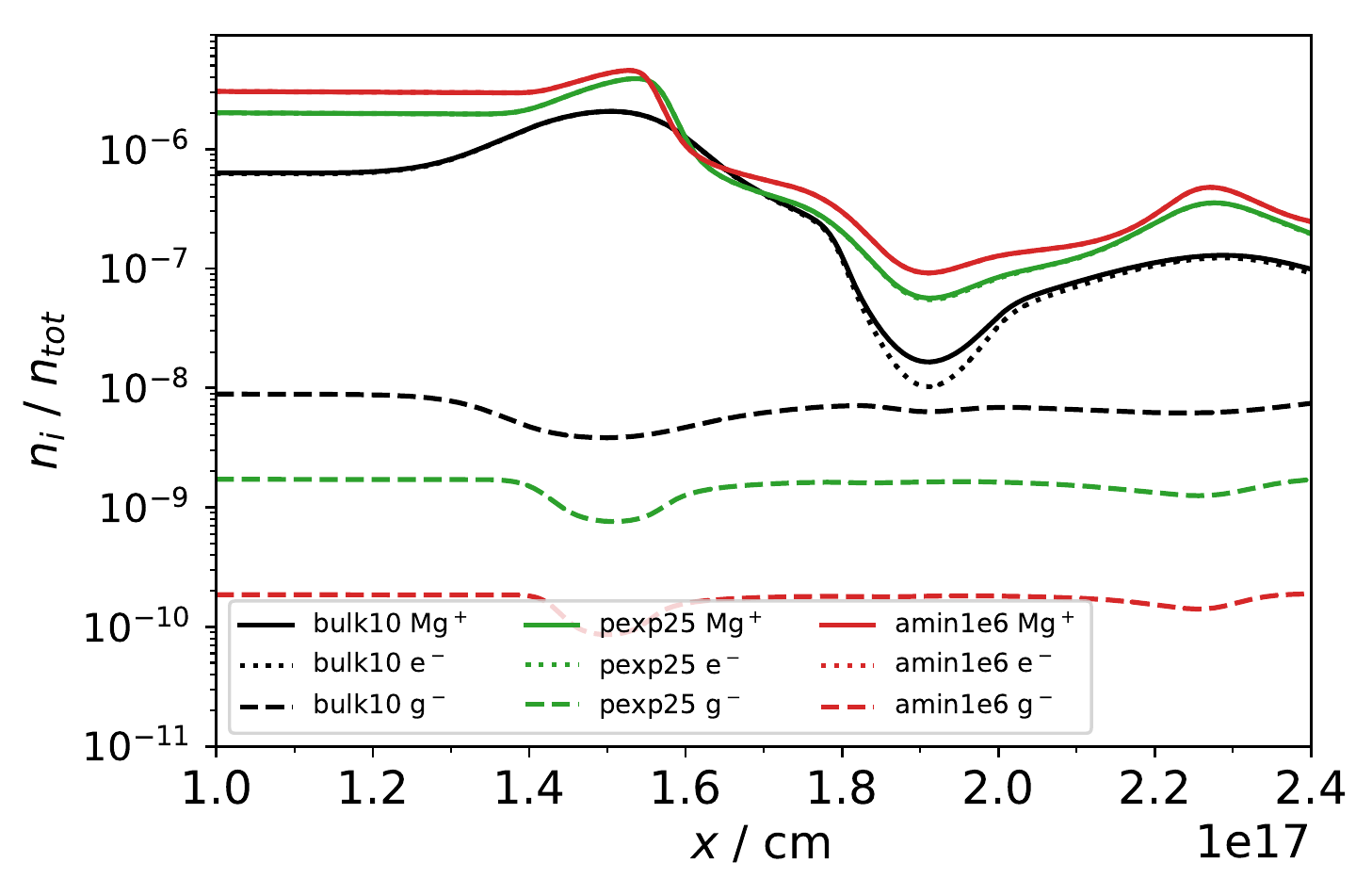}
 \caption{Number density fractions $n_i/n_{\rm tot}$ for electrons (dotted), Mg$^+$ (solid), and negatively charged grains (dashed) for different models, as listed in the legend.}
 \label{fig:relative_abundances}
\end{figure}

Figure \ref{fig:max_eAD} shows the contributions of Mg$^+$, e$^-$, and g$^-$ to $\eta_{\rm AD}$ in the \test{reference}, \test{stick01} ($S(T) = 0.1$), and \test{stick1} ($S(T) = 1$) models. Note that, since the fitting function for $S(T)$ (\sect{sect:sticking}) is of order 0.1 (see Fig.\ 6 in \citealt{Bai2011}), the \test{stick01} model is very similar to the \test{reference} model. An electron-grain sticking coefficient of $S(T) = 1$, however, results in a decreased abundance of e$^-$ and subsequent increased abundance of g$^-$. As can be seen in \fig{fig:relative_abundances}, the increase of g$^-$ sticking partners reduces the abundance of Mg$^+$, but because electron-grain sticking is more prevalent than cation-grain sticking, there is still a net increase in g$^-$ grains.  As \fig{fig:max_eAD} shows, negatively-charged grains are the most important contributor to $\eta_{\rm AD}$. Thus, increasing the sticking coefficient decreases the magnetic resistivity and moves the system towards the \test{ideal} case (Figs.\ \ref{fig:coloramagram_rho} -- \ref{fig:coloramagram_eAD}).

It is clear that the sticking coefficient plays an important role in determining the strength of the non-ideal terms, but we point out that its modelling is subject to significant uncertainties, such as in the depth of the potential well $D$ (see \sect{sect:sticking}). For example, in the cases reported in \citet{Bai2011}, and for the temperature range of the present study, the sticking coefficient has values in the range $0.1\lesssim S(T) \lesssim 0.7$ for $1\leq D\leq 3$~eV.

\subsubsection{Constant ionisation fraction}
The ionisation fraction is determined by the chemistry. Thus, if we turn off chemistry altogether and instead force a constant ionisation fraction, we naturally find quite different results relative to the \test{reference} model.

In models \test{nochem1e4}, \test{nochem1e5}, \test{nochem1e6}, and \test{nochem1e7} the ionisation fraction is set to a constant $f_i=10^{-4}$ to $10^{-7}$, respectively. In these cases, the chemical initial conditions remain unaltered during the evolution, which means $n_{\rm g^\pm}=0$, $n_{\rm e^-}=n_{\rm Mg^+}=f_i n_{\rm H_2}$ (see \appx{sect:initial}); grains remain neutral, and electrons and Mg$^+$ never recombine. 

The evolution and structure of the shock is therefore controlled by $f_i$ and, when the ionisation fraction is large (e.g.\ \test{nochem1e4}), the results approach the \test{ideal} MHD limit. Conversely, when it is low (e.g.\ \test{nochem1e7}), the magnetic field is strongly diffused and nearly passive (\fig{fig:coloramagram_B}). In fact, the models with constant ionisation fractions of $10^{-4}$ and $10^{-7}$ produce the largest deviations from the \test{reference} model and set the bounds of the gray envelopes in Figs.\ \fig{fig:coloramagram_rho} -- \fig{fig:coloramagram_eAD}.

Evidently, a consistently calculated ionisation fraction is critical for obtaining a physically realistic shock structure and evolution \citep[e.g.][]{Flower1985}. While it is true that the \test{nochem1e6} model results are relatively similar to the \test{reference} model, this is only known because we first did the calculation consistently. As such, we caution others from using constant ionisation fractions when calculating non-ideal coefficients, unless it has already been established with a full calculation that this is a good approximation.

\subsection{The effect of the cosmic ray parameters}
In the present set up, since we do not include any external radiation, cosmic rays are the main driver of ionisation, and their effect is therefore of paramount importance. This is similar to the conditions at high column depths in a quiescent molecular cloud with little ongoing star formation.

\subsubsection{Initial effective column density}
The first parameter we modify is the initial effective column density $N_{\rm eff,1}$, i.e, the assumed column density that the cosmic rays have travelled through before entering the simulation box at $x=0$ (see \fig{fig:shock_sketch}). The default value is $N_{\rm eff,1}=5\times10^{21}$~cm$^{-2}$, which is comparable to the accumulated column density from $x=0$ to $x\simeq2\times10^{17}$~cm (i.e.\ the position of the shock front at $t=10^4$~yr); given $n_{\rm tot}\simeq10^4$~cm$^{-3}$ (the average density of the shocked gas), $N_{\rm shock}=n_{\rm tot}\cdot x\simeq2\times10^{21}$~cm$^{-2}$ (Eq.\ \ref{eqn:sum_neff}). If we instead choose $N_{\rm eff,1}=5\times10^{20}$~cm$^{-2}$ (model \test{N5e20}), $N_{\rm shock}$ now dominates over $N_{\rm eff,1}$. Conversely, if we choose $N_{\rm eff,1}=5\times10^{22}$~cm$^{-2}$ (model \test{N5e22}), the opposite is true. The relative importance of $N_{\rm eff,1}$ with respect to $N_{\rm shock}$ explains the behaviour of the \test{N5e20} and \test{N5e22} models in Figs.\ \ref{fig:coloramagram_rho}--\ref{fig:coloramagram_eAD}. In the first case, $N_{\rm shock}$ dominates, so the variation of column density along the shock is relevant, even more than it is for the default value ($N_{\rm eff,1}=5\times10^{21}$~cm$^{-2}$; \test{reference} model). In the second case, $N_{\rm eff,1}$ dominates and $N_{\rm shock}$ is nearly ignorable. This is also clear from \fig{fig:zeta}, where the ionisation rate in the \test{N5e22} model is almost constant throughout the simulation box. Indeed, in Figs.\ \ref{fig:coloramagram_rho}--\ref{fig:coloramagram_eAD} and \fig{fig:zeta}, the \test{N5e22} model is most similar to the \test{cr16} model, which employs a constant cosmic ray ionisation rate of $\zeta=10^{-16}$~s$^{-1}$ (also see below).

\subsubsection{Constant ionisation and a ``low'' spectrum}
\begin{figure}
   \centering
       \includegraphics[width=.48\textwidth]{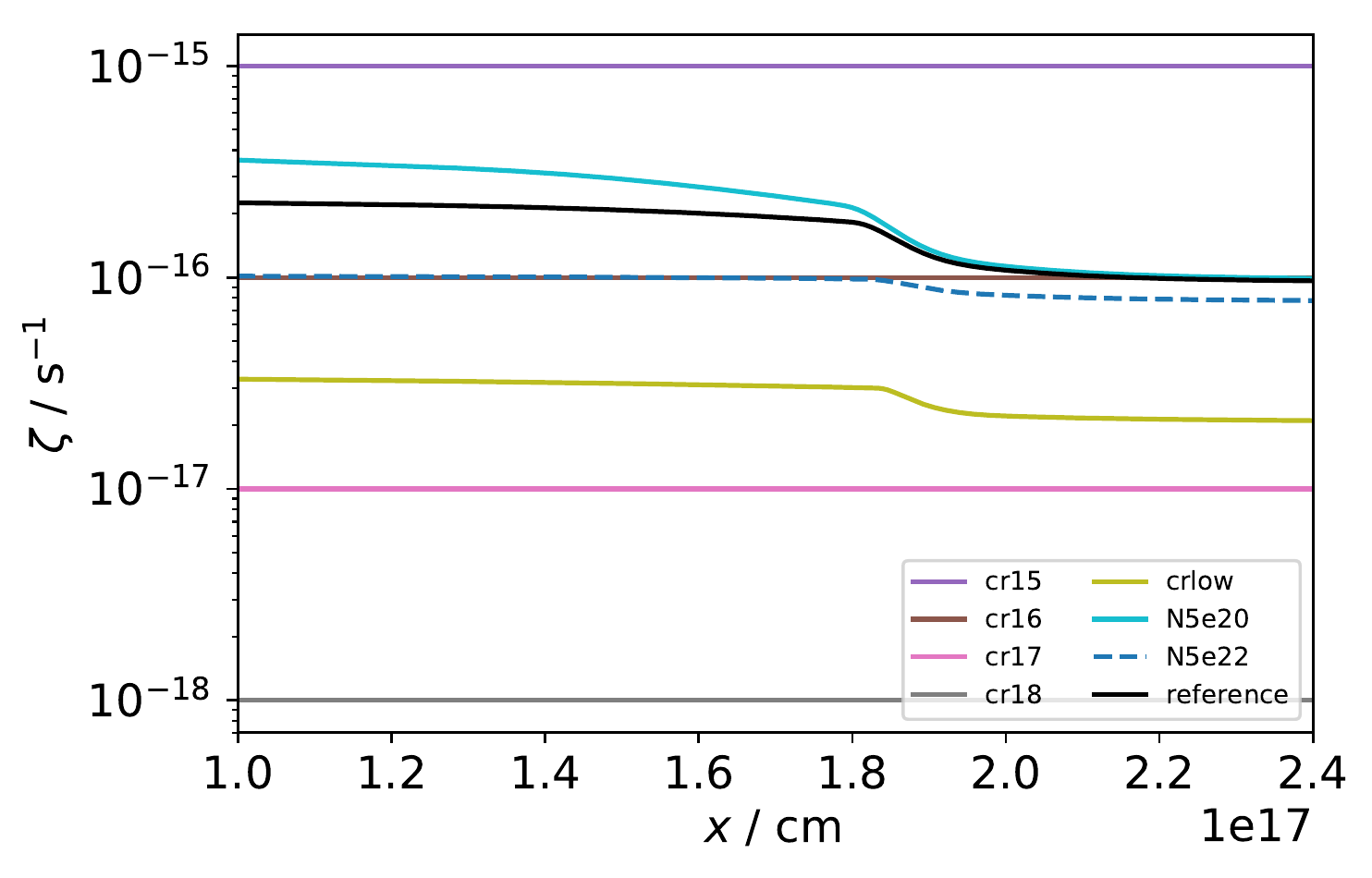}
 \caption{Cosmic ray ionisation rate at $t = 10^4$~yr as a function of $x$ for the cosmic ray related models listed in the legend and described in \tab{tab:tests_report}.}
 \label{fig:zeta}
\end{figure}

Next, we evolved shock models with a set of constant ionisation rates, using $\zeta=10^{-18}, 10^{-17}, 10^{-16}$, and $10^{-15}$~s$^{-1}$, and labelled with \test{cr}$x$ indicating a fixed $\zeta=10^{-x}$~s$^{-1}$. In general, since a larger $\zeta$ generates a larger number of free electrons, and thus a higher ionisation rate, the closer to the \test{ideal} model the results will be. Analogously, when we use the ``low'' spectrum fit (model \test{crlow}), i.e., \eqn{eqn:cr_fit} with $b_1=1.327\times10^{-12}$~s$^{-1}$ and $b_2=-0.211$, we observe greater non-ideal behaviour relative to the \test{reference} model. We also observe that the \test{crlow} model lies between \test{cr16} and \test{cr17} (\fig{fig:zeta}).

Evidently, the treatment of cosmic ray attenuation and subsequent ionisation rate plays a role in the shock structure and evolution. In particular, because cosmic rays control the ionisation fraction in our set up, they play an important role in determining the resistivity coefficients. Meanwhile, from \fig{fig:zeta}, it is clear that, when including cosmic ray attenuation consistently, it is difficult to obtain an ionisation rate comparable to the canonical, constant rate of $10^{-17}$~s$^{-1}$ \citep{Spitzer1968}, even if there is a substantial attenuating column between a region of interest and the source of cosmic rays. Even if one adopts a ``low'' energy CR spectrum (see \sect{sect:cosmic_rays_LH}), the resultant ionisation rate is still a factor of 2-3 above the canonical ionisation rate.

\subsection{The effect of different dust parameters}

\subsubsection{Dust-to-gas mass ratio}
As we already know from \sect{sect:vary_sticking}, charged grains play a key role in determining the non-ideal behaviour of the shock. Thus, if we decrease the dust-to-gas ratio from $\mathcal{D}=10^{-2}$ to $10^{-4}$ (model \test{d2g1e4}), we expect that the contribution to $\eta_{\rm AD}$ from (negative) grains will decrease. A reduction in the amount of dust will also naturally decrease the probability that Mg$^+$ and e$^-$ stick to grains. These effects are shown in \fig{fig:max_eAD}, where it can be seen that decreasing the dust-to-gas ratio produces a large decrease in $\eta_{\rm AD}$. In this case, $\eta_{\rm AD}$ is dominated by Mg$^+$ while the contribution from g$^-$ grains is negligible. Consequently, the shock structure is similar to the \test{ideal} case.

Next, we consider a discontinuous dust-to-gas ratio with a very low value ($\mathcal{D}=10^{-6}$) in the initially upstream region (i.e.\ $x < 0.3L_{\rm box}$), but a typical value ($\mathcal{D}=10^{-2}$) in the initially downstream region. This model is intended to mimic the propagation of a dust-free shock into a dense, cold, and dust-rich cloud. The upstream region of the \test{d2g\_step} model ($x\lesssim 1.7\times 10^{17}$~cm; \fig{fig:max_eAD}) is very similar to the same region in the \test{d2g1e4} model. In contrast, the downstream region ($x\gtrsim 1.9\times 10^{17}$~cm) is similar to the \test{reference} model, where $\mathcal{D}=10^{-2}$. This behaviour suggests that the advection of dust (or lack thereof) can strongly affect $\eta_{\rm AD}$ and the shock structure. However, the partial overlap between \test{d2g1e4} and \test{reference} models with respect to \test{d2g\_step} indicates that the net effect is less trivial than the sum of two dust-to-gas ratios.

\subsubsection{Grain size distribution}
To understand the effect of the grain size distribution $\varphi\propto a^p$ on the shock evolution, we now modify\footnote{We do not vary $a_{\rm max}$, since with a power-law distribution the smaller grains comprise the largest total surface.} $a_{\rm min}$, $p$, and the bulk density $\rho_0$. The impact of these parameters on $\eta_{\rm AD}$ is reported in \fig{fig:max_eAD}. First, in \test{amin1e6}, we adopt a larger minimum grain size ($a_{\rm min}=10^{-6}$~cm) relative to the \test{reference} model ($a_{\rm min}=10^{-7}$~cm). With the removal of smaller grains, similar to \test{d2g1e4}, $\eta_{\rm AD}$ decreases and is now dominated by Mg$^+$.

Analogously, in \test{pexp25}, if we adopt a shallower power-law exponent ($p=-2.5$; the \test{reference} model has $p=-3.5$), the dust mass becomes more heavily distributed towards larger grains, with a net effect on $\eta_{\rm AD}$ that is similar, but slightly less evident, than in \test{amin1e6}. Evidently, small grains dominate $\eta_{\rm AD}$ and, in both cases, their removal results in a solution that tends toward the \test{ideal} case (albeit, not as strongly as in \test{d2g1e4}).

If we now adopt a steeper power-law exponent, $p=-5$ (\test{pexp50}), the resulting shift of dust mass to smaller sizes decreases the abundances of e$^-$ and Mg$^+$ while enhancing the fraction of negatively charged grains (see, e.g., \test{stick1}; \sect{sect:vary_sticking}). This is visible in \fig{fig:relative_abundances}, where the relative abundance of Mg$^+$ decreases relative to the \test{reference}, and becomes comparable to the fraction of g$^-$ grains. The effect of the steeper power-law is thus a smaller overall value of $\eta_{\rm AD}$ (similar to \test{stick1}), but still higher than the \test{pexp25} and \test{amin1e6} models.

Finally, by increasing the bulk grain density from 3 to 10 g~cm$^{-3}$ (model \test{bulk10}), we find that $\eta_{\rm AD}$ is somewhat reduced (relative to \test{reference} and similar to \test{pexp50}), but dominated by Mg$^+$ instead of g$^-$ grains (\fig{fig:max_eAD}). Since we are keeping the total dust mass constant, increasing the bulk density effectively decreases the number of grains, which results in less recombinations of cations and electrons with charged grains, and therefore larger abundances of these species (\fig{fig:relative_abundances}).

\subsubsection{Grain-grain reactions}
In the \test{nogg} model, we examine the effect of grain-grain reactions by removing them from the network. As can be seen in Figs.\ \ref{fig:coloramagram_rho}--\ref{fig:coloramagram_eAD}, \ref{fig:max_eAD}, the effects on the shock structure and evolution are negligible. In the current context, grain-grain reactions are only marginally involved; the chemistry is dominated by the production of Mg$^+$ and e$^-$ from H$_2$ ionisation via cosmic rays, followed by the interaction of electrons and cations with grains to form g$^-$ and g$^+$, which then eventually recombine with Mg$^+$ to form H$_2$. Similarly, as was also shown in M16, grains with charge $Z=\pm2$ do not play a key role in the chemistry.

\subsection{The effect of changing the thermal processes}\label{sect:vary_thermal}
\begin{figure}
   \centering
       \includegraphics[width=.5\textwidth]{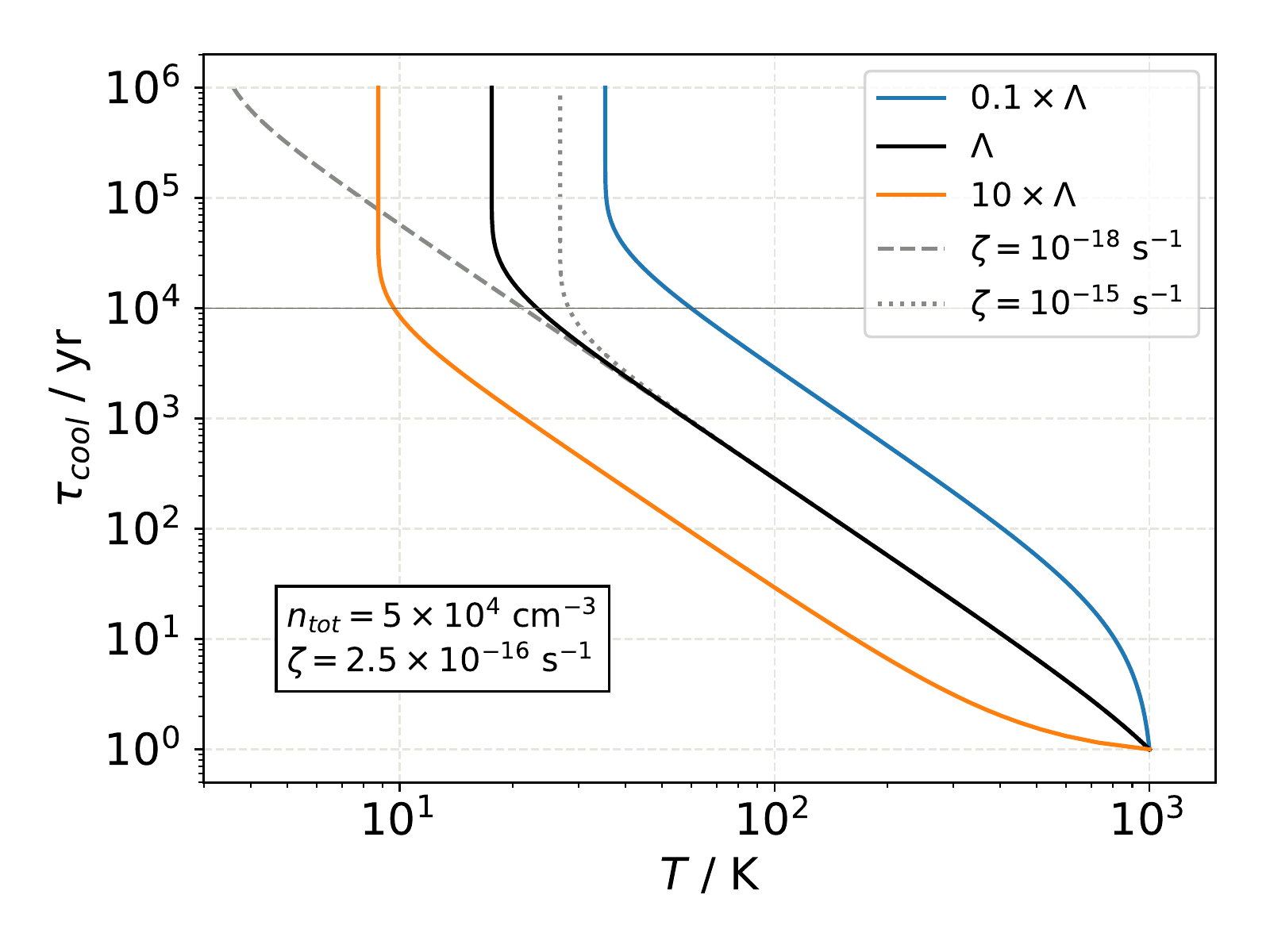}
 \caption{The cooling time scale $\tau_{\rm cool}$ as a function of the temperature, assuming $n_{\rm tot}=10^4$~cm$^{-3}$, $\zeta=2.5\times10^{-16}$~s$^{-1}$, and initial temperature $10^3$~K. For reference, the dashed line assumes $\zeta=10^{-18}$~s$^{-1}$ in $\Gamma_{\rm CR}$, while the dotted assumes $\zeta=10^{-15}$~s$^{-1}$. The solid, horizontal line represents $\tau_{cool}=10^4$~yr, the evolution time of the shock. Cooling is decreased (label $0.1\times\Lambda$) or increased by a factor of ten ($10\times\Lambda$) with respect to the \test{reference} ($\Lambda$). Note that ambipolar diffusion heating is not included in the calculation of the cooling time scale.}
 \label{fig:ctime}
\end{figure}

To better understand the temperature structure of the shock (\fig{fig:coloramagram_T}), we compute the cooling time by combining Eqs.\ (\ref{eqn:pressure}) and (\ref{eqn:pressure_temperature}) and taking the time derivative (see \appx{sect:cooling_time} for details):
\begin{equation}\label{eqn:cooling_time_ode}
 \frac{\dd T}{\dd t}=(\gamma-1) \frac{\Gamma_{\rm CR}-\Lambda(T)}{n_{\rm tot} k_{\rm B}}\,.
\end{equation}
We integrate this equation numerically\footnote{Using the \textsc{odeint} solver from the \textsc{scipy} package.} assuming typical values $n_{\rm tot}=10^4$~cm$^{-3}$, $\zeta=2.5\times10^{-16}$~s$^{-1}$, and initial temperature $10^3$~K. The results are shown in \fig{fig:ctime}.

We note that the cooling time $\tau_{\rm cool}$ follows $\Lambda\propto T^{3.3}$ (Eq.\ \ref{eqn:chemical_cooling}) and, given that we evolve the system for $t=10^4$~yr, it is not surprising that the gas decreases to $T < 100$~K. The gas does not have time to cool further, however, and thus does not reach the equilibrium temperature where $\Lambda=\Gamma_{CR}$ (which corresponds to a vertical line in \fig{fig:ctime}, i.e., $\tau_{\rm cool}\to\infty$). If we now reduce the cooling function by a factor of 10 ($0.1\times\Lambda$ in \fig{fig:ctime}), the cooling time increases, and the temperature in the shock and upstream region increases (see also \fig{fig:coloramagram_T}). Analogously, if we increase the cooling by a factor of 10 ($10\times\Lambda$ in \fig{fig:ctime}), the gas naturally reaches lower temperatures ($T<10$~K) in a shorter time.

This trend is confirmed in \fig{fig:coloramagram_T}, where, for the \test{cool01} model, the temperature is higher relative to the \test{reference} everywhere but in the quiescent downstream region. Indeed, the \test{cool01} model produces the highest temperatures, in general, of all the models examined. For \test{cool10}, meanwhile, the opposite is true and, because the cooling time scale is now shorter, the quiescent downstream region even cools somewhat with respect to the initial conditions.

If we instead examine the effect of the cosmic ray heating rate $\Gamma_{\rm CR}$ on the cooling time, we find that the equilibrium temperature correlates with the ionisation rate $\zeta$. While it does not affect the cooling time scale where $\Lambda$ dominates, as can be seen in \fig{fig:ctime}, an increased cosmic ray ionisation rate of $\zeta=10^{-15}$~s$^{-1}$ does raise the minimum temperature slightly, while decreasing the value to $\zeta=10^{-18}$~s$^{-1}$ decreases the equilibrium temperature to $T < 4$~K.

That said, when the cosmic ray heating term is turned off entirely (model \test{noheat}), the impact on the temperature profile (see \fig{fig:coloramagram_T}) is negligible, in contrast to what one would expect from \fig{fig:ctime}. The temperature in models \test{cr18} and \test{cr15} (with constant cosmic ray ionisation rates of $\zeta=10^{-18}$~s$^{-1}$ and $10^{-15}$~s$^{-1}$, respectively) does, meanwhile, deviate from the \test{reference}. In fact, the effect of a constant cosmic ray ionisation rate on the temperature is not direct via cosmic ray heating but rather indirectly through the chemistry and MHD heating processes.

\begin{figure}
   \centering
       \includegraphics[width=.48\textwidth]{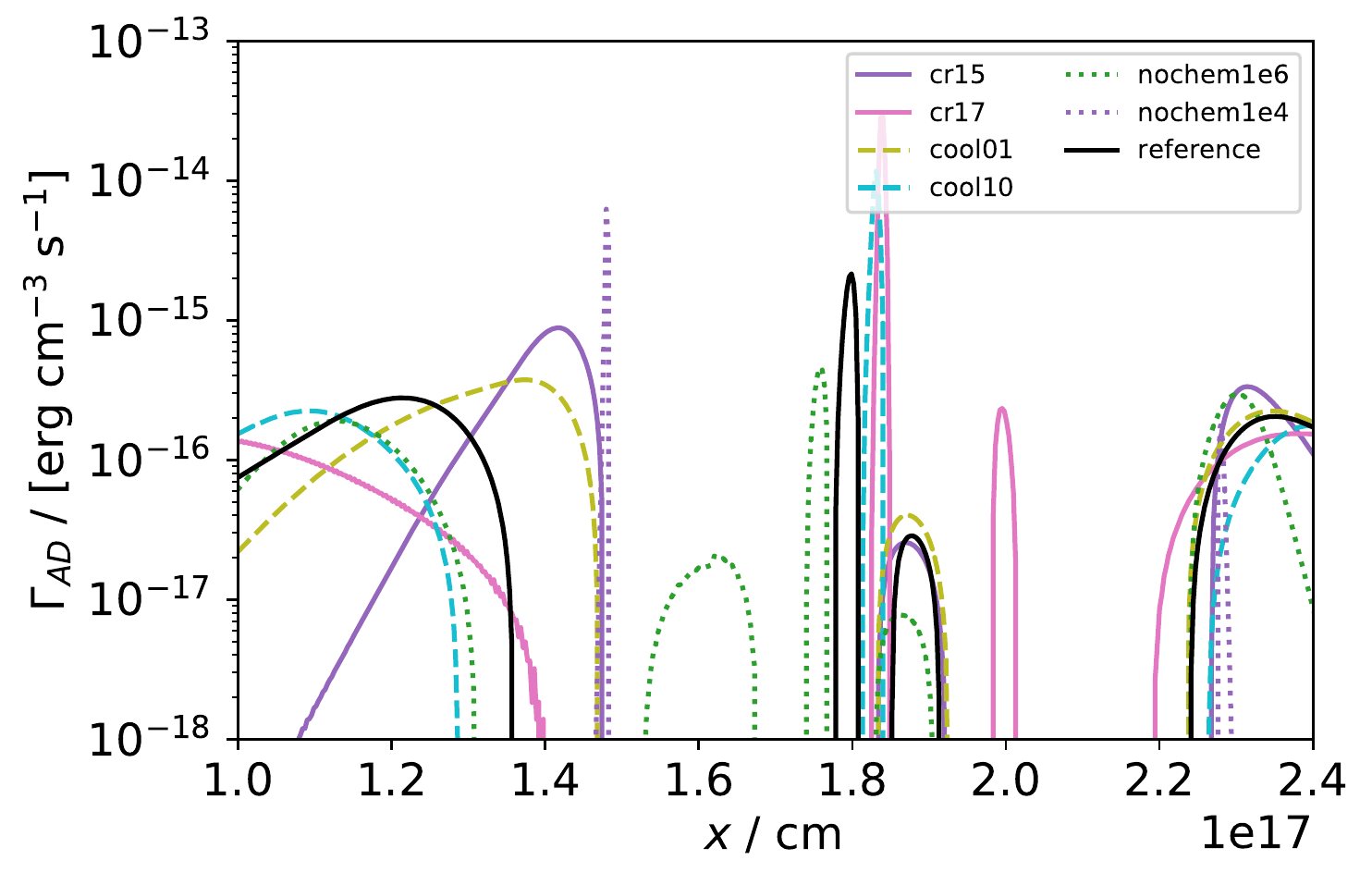}
 \caption{Ambipolar diffusion heating rate $\Gamma_{\rm AD}$ as a function of $x$ for select models at $t = 10^4$~yr, as compared to the \test{reference} model (solid black).}
 \label{fig:ad_heating}
\end{figure}

The temperature behaviour in the different models suggest that it is primarily the ambipolar diffusion heating that prevents the temperature from reaching the levels observed in the \test{ideal} MHD model, which indeed provide the lowest temperatures observed in any of the models considered here. To demonstrate this, in \fig{fig:ad_heating}, we plot the the ambipolar diffusion heating rate, which is given by the spatial derivative of the third term of \eqn{eqn:hydro_full_energy}. Since $\partial_x B_x=0$ and $B_z=0$, the heating rate can be written as
\begin{equation}\label{eqn:ad_heating_rate}
 \Gamma_{\rm AD}(x) = \frac{\eta_{\rm AD}}{4\pi}\frac{\partial^2B_y(x)}{\partial x^2}\,.
\end{equation}
Comparing the ambipolar heating rate (\fig{fig:ad_heating}) with the temperature (\fig{fig:coloramagram_T}), it is clear that the trends observed in the temperature of the different models can be explained by heating due to ambipolar diffusion.

\section{Conclusions}\label{sect:conclusions}
We have developed an open-source\footnote{\linktocode} 1D, time-implicit, MHD code that includes ambipolar diffusion, chemistry, dust, and consistent cosmic-ray propagation. The code has been employed to explore the evolution of an oblique magnetic shock in order to understand the effects of the different microphysical parameters on the results. We have shown that, even in a simple application, microphysics plays a crucial role, and that the uncertainties are manifold.

Our main findings can be summarised as follows:
\begin{itemize}[topsep=0pt]
\item In the absence of external radiation, cosmic rays plays a key role by controlling the ionisation level of the gas (via the chemistry) and hence determining the amplitude of the non-ideal MHD effects. In this study, we find that the widely-used cosmic-ray ionisation rate of $\zeta=10^{-17}$~s$^{-1}$ results in up to $\sim90\%$ relative error in the density and $\sim60\%$ in the magnetic field strength after $10^4$~yr relative to a self-consistent treatment of propagation.

\item Dust is responsible for a large fraction of the neutral-ion momentum exchange. Reducing the dust-to-gas mass ratio from $\mathcal{D}=10^{-2}$ to $\mathcal{D}=10^{-4}$ results in a $\sim280\%$ relative change in the density and a $\sim60\%$ relative change in the magnetic field.

\item The gas density is strongly affected by the parameters of the grain size distribution. Increasing the lower size limit of the distribution to $10^{-6}$~cm from $10^{-7}$~cm produces a change of $\sim200\%$ in the density.

\item Chemistry is also paramount. When turned off, the ionisation fraction is arbitrary (and constant in time), which affects the resistivity. A low ionisation fraction of $f_i=10^{-7}$ results in very strong ambipolar diffusion while, in contrast, values of $f_i=10^{-5}$ and $10^{-4}$ produce results that are very similar to the ideal MHD case.

\item Reducing the cosmic-ray ionisation rate increases the temperature in certain regions, not because the direct cosmic-ray heating decreases, but because the ambipolar diffusion heating increases in these regions as a result of the lower ionisation. Analogously, higher ionisation rates (i.e.\ more highly ionised gas) show lower temperatures. This effect could be limited to the present set up, but due to its potential consequences, is worth exploring in more complex 3D simulations of pre-stellar cores and other environments.

\item We find almost no change in the results when direct cosmic-ray heating is turned off or when grain-grain reactions are removed.
\end{itemize}

We remark that, given the complexity of the processes discussed in this paper and the interactions between them, our findings are particularly valid within the current set-up, and \emph{should not be arbitrarily generalised}. For example, models with higher dimensions, complex chemistry, more realistic cooling functions, or other additional physics could change the importance of the different physical processes. To understand the interplay between the physical processes, the chemical model employed in this study was intentionally kept simple; determining the effects of a network with thousands of reactions, where many rates coefficients are uncertain, is beyond the aims of this paper. Nevertheless, the tests presented here show that self-consistent microphysics \emph{cannot be ignored in the context of non-ideal MHD simulations}, and the choice of processes and parameters (mainly chemistry, cosmic rays, and dust) significantly affects the evolution of the dynamics.

\section*{Acknowledgement}
We acknowledge Y.~Fujii, O.~Gressel, C.~McNally, and R.~Xu for useful comments and discussions.
MP acknowledges funding from the European Unions Horizon 2020 research and innovation programme under the Marie Sk\l{}odowska-Curie grant agreement No 664931.
The research leading to these results has received funding from the Danish Council for Independent Research through a Sapere Aude Starting Grant to TH. The Centre for Star and Planet Formation is funded by the Danish National Research Foundation (DNRF97).
This research was supported by the DFG cluster of excellence ``Origin and Structure of the Universe''
(\url{http://www.universe-cluster.de/}).
This work was funded by the Deutsche Forschungsgemeinschaft (DFG, German Research Foundation) - Ref no. FOR 2634/1 ER685/11-1.
We would also like to thank the referee T.~Hartquist for bringing additional relevant references to our attention.

\bibliographystyle{mn2e}
\bibliography{mybib}

\appendix
\section{Algorithm testing}\label{sect:tests}
To validate the code implementation, we present here results for two standard MHD benchmarks. First, a Brio-Wu shock tube \citep{Brio1988} with ideal MHD and, second, a C-shock tube \citep{Masson2012} with MHD and ambipolar diffusion. 

\subsection{Brio-Wu MHD shock}
\begin{figure}
  \centering
  \includegraphics[width=.5\textwidth]{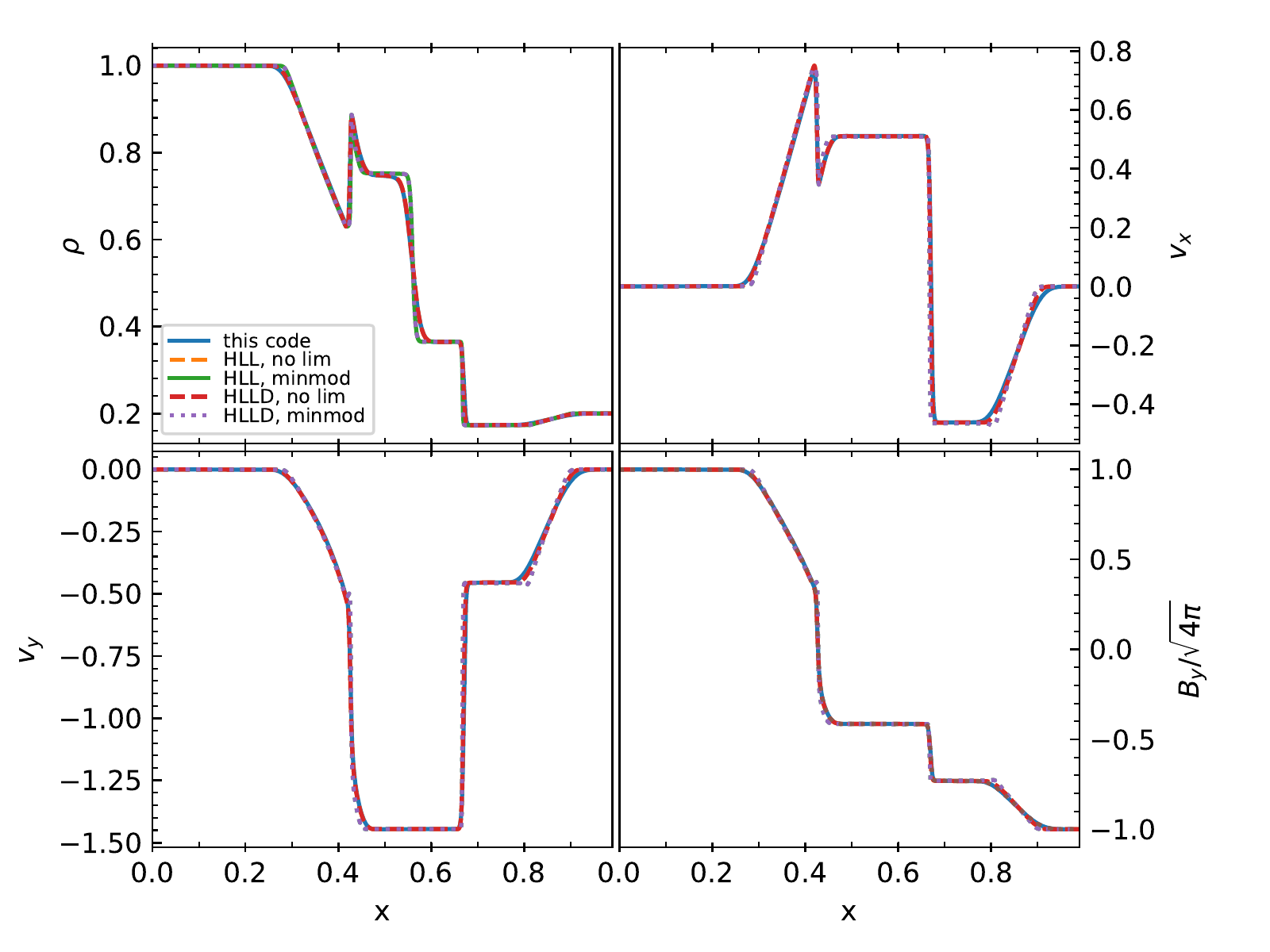}
  \caption{Comparison of results between our HLL solver and the HLL/HLLD solvers in \ramses{} with and without \textsc{MinMod} slope limiter for the Brio-Wu shock test. Plotted from left to right, top to bottom, are the density, $x$- and $y$- components of velocity, and $y$-component of the magnetic field at $t=0.12$. Units are arbitrary. Note that the results overlap with the exception that the solver with \textsc{MinMod} is slightly less diffusive.}
  \label{fig:brio_wu}
\end{figure}

To test the (ideal) MHD, we performed a Brio-Wu shock test \citep{Brio1988} with $\mathbf{U}=\mathbf{0}$, except for $(\rho, P, B_y)=(1, 1, \sqrt{4\pi})$ where $x\le0.5$ and $(\rho, P, B_y)=(0.2, 0.2, -\sqrt{4\pi})$ otherwise, $B_x=\sqrt{4\pi}$, $\gamma=5/3$, $x\in[0,1]$. We used 1024 grid points. Note that for \ramses{} we employed rational units, i.e.~$B_{\rm ramses} = B/\sqrt{4\pi}$. In \fig{fig:brio_wu}, at $t=0.12$, we present our results. We also compared our HLL solver with the \ramses{} HLL and HLLD solvers, respectively with no slope limiter (\texttt{slope\_type=0}) and \textsc{MinMod} limiter (\texttt{slope\_type=1}) \citep{Teyssier2002,Fromang2006,Teyssier2006}. As expected, the solver without slope limiter is slightly more diffusive. We note that our results are indistinguishable from those obtained with the \ramses{} solver without slope limiter.

\subsection{Non-ideal MHD C-shock}
\begin{figure}
   \centering
       \includegraphics[width=.5\textwidth]{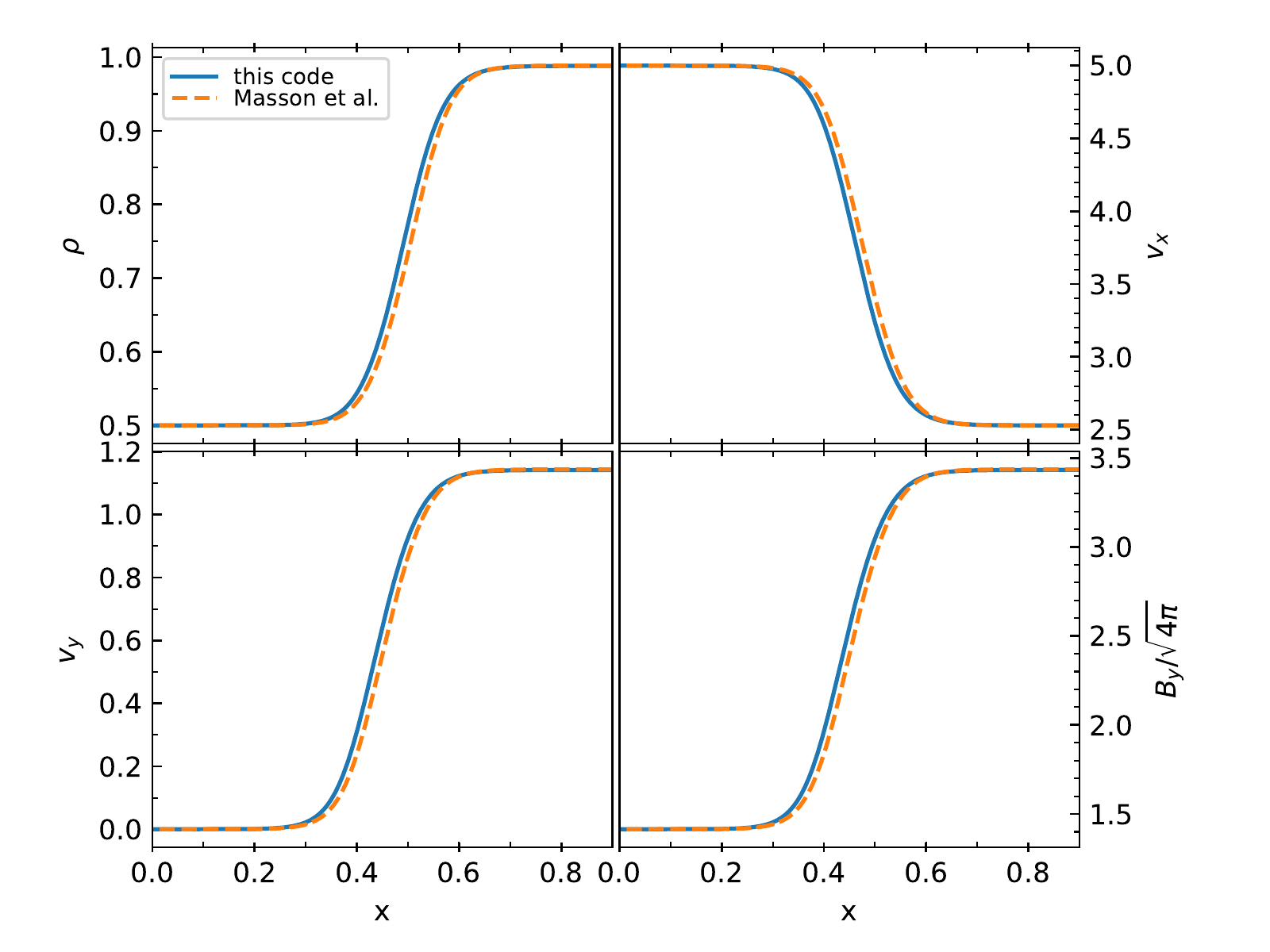}
 \caption{Comparison of our results between our code and \citet{Masson2012} for a non-isothermal C-shock. Plotted from left to right and top to bottom are the density, $x$- and $y$-components of the velocity, and $y$-component of the magnetic field. As can be seen, at $t=1$, our results agree sufficiently well with the analytical solution from \citet{Masson2012}. Units are arbitrary.}
 \label{fig:masson}
\end{figure}

To determine if our ambipolar diffusion implementation is functioning properly, we benchmarked our code against the non-ideal MHD, non-isothermal C-shock presented in \citet[][Sect.\ 2.4.2]{Masson2012}. The initial conditions are $\mathbf{U}=\mathbf{0}$, except for $(\rho, v_x, v_y, P, B_y)=(0.5, 5, 0, 0.125, \sqrt{2})$ where $x\le0.5$ and $(\rho, v_x, v_y, P, B_y)=(0.988, 2.5303, 1.1415, 1.4075, 3.4327)$ otherwise, $B_x=\sqrt{2}$, $\gamma=5/3$, $\gamma_{\rm AD}=75$, $\rho_{\rm ion}=1$, $x\in[0,1]$. We used 400 grid points. Note that \citet{Masson2012} use rational units, hence $B_{\rm Masson} = B/\sqrt{4\pi}$. We evolve the system until the shock reaches a stationary state. In \fig{fig:masson}, we compare our results at $t=1$ to the analytic solution of \citet{Masson2012}. As can be seen, other than a very slight difference in the position of the shock, we accurately reproduce the analytic solution.

\section{Grain size distribution properties}\label{sect:grain_distribution}
In this study, the grain size distribution is given by $\varphi(a) \propto a^p$ between sizes $a_{\rm min}$ and $a_{\rm max}$. The average grain mass is thus
\begin{equation}
 \langle m_{\rm d} \rangle = \frac{4 \pi\rho_0 }{3}\cdot \frac{\int_{a_{\rm min}}^{a_{\rm max}} \varphi(a) a^3 \dd a}{\int_{a_{\rm min}}^{a_{\rm max}} \varphi(a) \,\dd a}\,,
\end{equation}
and from whence we compute the corresponding dust number density needed by the chemistry:
\begin{equation}
 n_{\rm d} = \frac{\rho_{\rm d}}{\langle m_{\rm d} \rangle} = \frac{\rho \mathcal{D}}{\langle m_{\rm d} \rangle} = \frac{n_{\rm g}\mu_{\rm g} m_{\rm p} \mathcal{D}}{\langle m_{\rm d} \rangle}\,,
\end{equation}
where $\rho$ and $\rho_{\rm d}$ are the mass densities of gas and dust, respectively, 
$\mu_g$ is the dust mean molecular weight, and $\mathcal{D}$ is the dust-to-gas mass ratio.

To better compare our results with models that employ a constant grain size, we also derive the average surface area
\begin{equation}\label{eqn:average_area}
 \langle a^2 \rangle = \frac{\int_{a_{\rm min}}^{a_{\rm max}} \varphi(a) a^2 \dd a}{\int_{a_{\rm min}}^{a_{\rm max}} \varphi(a) \,\dd a}\,.
\end{equation}
With $p=-3.5$, and $a_{\rm min}=10^{-7}$ to $a_{\rm max}=10^{-5}$~cm, this corresponds to an average grain size of $\sqrt{\langle a^2 \rangle} = 2.1\times10^{-7}$~cm.

\section{Polynomial fitting functions for grain reactions}\label{sect:k_poly_grains}
Here, we report the fitting functions for the reaction rates that involve dust grains discussed in \sect{sect:grain_chemistry}. The grain distribution used here is $\varphi(a)\propto a^{p}$ with $p=-3.5$ and $a_{\rm min}=10^{-7}$ to $a_{\rm max}=10^{-5}$~cm.  The fitting functions are of the form
\begin{equation}\label{eqn:fit_grain_reactions}
	\log[k(T)] = \sum_{i=0}^5 c_i \log(T)^i\,,
\end{equation}
with $T$ in~K, $k(T)$ in~cm$^{3}$~s$^{-1}$, and coefficients $c_i$ given in \tab{table:grain_rates_fit}. Fits are valid in the range $T=3$~K to $T=10^4$~K.

\begin{table*}
\begin{tabular}{lrrrrrr}
\hline
Reactants & $c_{0}$ & $c_{1}$ & $c_{2}$ & $c_{3}$ & $c_{4}$ & $c_{5}$\\
\hline
Mg$^+$ + g & $-7.418(0)$ & $2.331(-4)$ & $1.619(-2)$ & $-9.891(-3)$ & $4.072(-3)$ & $-3.404(-4)$\\
Mg$^+$ + g$^-$ & $-5.268(0)$ & $-5.931(-1)$ & $1.688(-1)$ & $-1.089(-1)$ & $2.762(-2)$ & $-2.041(-3)$\\
g$^+$ + g$^-$$^-$ & $-6.110(0)$ & $-5.995(-1)$ & $1.778(-1)$ & $-1.133(-1)$ & $2.844(-2)$ & $-2.091(-3)$\\
e$^-$ + g$^-$ & $-2.997(1)$ & $3.028(1)$ & $-1.622(1)$ & $4.445(0)$ & $-5.975(-1)$ & $3.122(-2)$\\
e$^-$ + g$^+$ & $-2.946(0)$ & $-5.931(-1)$ & $1.688(-1)$ & $-1.089(-1)$ & $2.762(-2)$ & $-2.041(-3)$\\
g + g$^+$$^+$ & $-8.310(0)$ & $4.465(-3)$ & $1.149(-2)$ & $-7.213(-3)$ & $3.670(-3)$ & $-3.233(-4)$\\
e$^-$ + g$^+$$^+$ & $-2.705(0)$ & $-6.448(-1)$ & $2.285(-1)$ & $-1.302(-1)$ & $2.942(-2)$ & $-2.013(-3)$\\
g$^+$ + g$^-$ & $-6.266(0)$ & $-5.478(-1)$ & $1.109(-1)$ & $-8.398(-2)$ & $2.403(-2)$ & $-1.883(-3)$\\
g$^+$$^+$ + g$^-$$^-$ & $-6.266(0)$ & $-5.478(-1)$ & $1.109(-1)$ & $-8.398(-2)$ & $2.403(-2)$ & $-1.883(-3)$\\
g$^+$$^+$ + g$^-$ & $-6.146(0)$ & $-5.890(-1)$ & $1.646(-1)$ & $-1.079(-1)$ & $2.770(-2)$ & $-2.062(-3)$\\
g + g$^-$$^-$ & $-8.310(0)$ & $4.465(-3)$ & $1.149(-2)$ & $-7.213(-3)$ & $3.670(-3)$ & $-3.233(-4)$\\
Mg$^+$ + g$^+$ & $-2.985(1)$ & $2.590(1)$ & $-1.336(1)$ & $3.574(0)$ & $-4.728(-1)$ & $2.443(-2)$\\
Mg$^+$ + g$^-$$^-$ & $-5.027(0)$ & $-6.449(-1)$ & $2.285(-1)$ & $-1.302(-1)$ & $2.941(-2)$ & $-2.013(-3)$\\
e$^-$ + g & $-5.096(0)$ & $2.384(-4)$ & $1.618(-2)$ & $-9.888(-3)$ & $4.071(-3)$ & $-3.403(-4)$\\
\hline
\end{tabular}\caption{Coefficients for fitting function \eqn{eqn:fit_grain_reactions} applied to reaction rates that involve dust grains (\sect{sect:grain_chemistry}). In the first column, for simplicity, we indicate only the reactants. Coefficients in the table use the notation $a(b) = a\times10^b$ and have units of cm$^3$~s$^{-1}$.}
\label{table:grain_rates_fit}
\end{table*}

\section{Evaluating the charged grain--H$_2$ momentum transfer rate}\label{sect:grain_H2_simple}
\eqn{eqn:charge_grain_H2} evaluated for $p=-3.5$, $a_{\rm min}=10^{-7}$~cm, $a_{\rm max}=10^{-5}$~cm, $\delta=1.3$, and $\alpha_{\rm pol}=8.06\times10^{-25}$~cm$^3$ is equal to
\begin{equation}
 \langle R_g(T,Z) \rangle = a_1\sqrt{T} + a_2 |Z|^{-0.125}T^{0.625}+a_3\sqrt{|Z|}\,,
\end{equation}
where $a_1=-2.7913\times10^{-10}$, $a_2=1.7065\times10^{-9}$, and $a_3=1.6369\times10^{-9}$. All coefficients are in cgs units and the resulting rate is in~${\rm cm^3\,s^{-1}}$.

\section{Momentum transfer collisions}\label{sect:pinto_collisions}
\begin{figure}
   \centering
       \includegraphics[width=.48\textwidth]{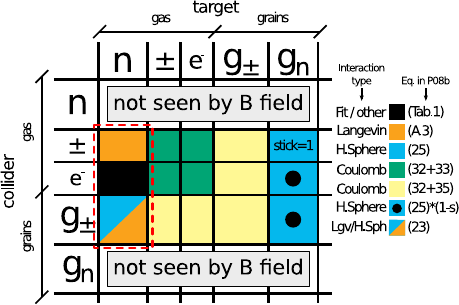}
 \caption{Reference table for the types of collisions described in \citet{Pinto2008b}. Colliders are listed on the left and targets on the top, where $n$ are neutral gas species, $\pm$ are ions, e$^-$ are electrons, $g_{\pm}$ are charged grains, and $g_{\rm n}$ are neutral grains. Colours correspond to the type of interaction and the equation number in P08b. A dashed box encloses the processes included in this work. Additional details can be found in \sect{sect:niMHD}.}
 \label{fig:collisions_table}
\end{figure}

In \fig{fig:collisions_table}, we report, for reference, the different processes described in \citet[][henceforth P08b]{Pinto2008b}. Colours indicate the type of interaction and the corresponding equation number in P08b. Magnetic fields only interact with neutrals indirectly via charged colliders. The processes which are included in this work are enclosed by a dashed box. The ``fit/other'' label denotes fits to experiments or theoretical calculations (see Tab.\ 1 of P08b); ``Langevin'' and ``Lgv'' refer to the Langevin model (see also our \sect{sect:charged_grain_H2}); ``H.Sphere'' and ``H.Sph'' refer to the hard sphere approximation and means the rate equations are multiplied by $(1-S)$ where $S$ is the sticking coefficient; ``Coulomb'' is the standard Coulomb momentum transfer rate and the equations employed in P08b for these interactions is indicated. We refer the reader to P08b and references therein for additional details.

\section{Additional details for the Marchand et al.\ (2016) benchmark}\label{appx:M16}
In this Appendix, we include additional details to aid the interested reader in reproducing the results of M16.

The chemical network employed by M16 is reported in their Tab.\ A1 and includes the same reactions as in UN90. For the sake of completeness, however, the network is presented in \tab{tab:UN90_network}. Here, we also report on their assumptions:
\begin{itemize}
	\item All \emph{molecular} ions except H$_3^+$ (i.e.\ O$_2^+$, HCO$^+$, OH$^+$, O$_2$H$^+$, CH$_2^+$, and CO$^+$) are represented by m$^+$.
	\item Metallic, atomic cations (excluding oxygen and carbon) are indicated with M$^+$.
	\item M and m are the neutral counterparts to M$^+$ and m$^+$, respectively.
	\item When O$^+$ is produced, it immediately turns into OH$^+$ (i.e.\ m$^+$) by reacting with H$_2$.
	\item Analogously, when H$_2^+$ is produced (due to cosmic ray ionisation; see their Tab.\ 2), it turns immediately into H$_3^+$.
	\item CO and C are the same species, since all neutral carbon is assumed to be in the form of CO.
	\item Neutral products are ignored, since their reservoir is assumed to be constant with time.
\end{itemize}
The network thus includes the following species: e$^-$, O, O$_2$, M, M$^+$, H$_2$, C, He, He$^+$, m, m$^+$, and H$_3^+$. Neutral (g$^0$) and charged dust (g$^\pm$) is also included. Given the assumptions listed above, we obtain \tab{tab:UN90_network}, the chemical network solved in M16. Note that, as written, mass is not conserved, but charge is.

The only difference between M16 and UN90 is given by the additional rates (A.1)-(A.3) in M16, which are taken from Eqs.\ (22)-(24) in \citet{Pneuman1965}, although the coefficients in those expressions are different from the ones reported by M16. Based on the code released by the authors\footnote{Commit 2b23528.}, these rates should in fact be
\begin{eqnarray}\label{eqn:M16_thermoionization}
	\frac{\dd n_{\rm K^+}}{\dd t} &=& 4.1\times10^{-15}n_{\rm H_2} n_{\rm K} \cdot \sqrt{\frac{T_{\rm gas}}{10^3}}\\
                        &&\times\exp\left(-\frac{5.04\times10^4\,{\rm K}}{T_{\rm gas}}\right)\nonumber,\\
	\frac{\dd n_{\rm Na^+}}{\dd t} &=& 2.8\times10^{-15}n_{\rm H_2} n_{\rm Na} \cdot \sqrt{T_{\rm gas}}\\
                        &&\times\exp\left(-\frac{6\times10^4\,{\rm K}}{T_{\rm gas}}\right),\\
	\frac{\dd n_{\rm H^+}}{\dd t} &=& 2\times10^{-10}n_{\rm H_2}^2 \cdot \sqrt{T_{\rm gas}}\\
                        &&\times\exp\left(-\frac{15.8\times10^4\,{\rm K}}{T_{\rm gas}}\right)\nonumber,
\end{eqnarray}
and are relevant when $T_{\rm gas}>10^4$~K and $n_{\rm tot}\gtrsim10^{15}$~cm$^{-3}$.

Dust evaporates as explained in Sect.\ 2.4.2 of M16 (see their Fig.\ 2); this behaviour can be reproduced by scaling the dust-to-gas mass ratio by a factor
\begin{equation}
	f_{\rm evap} = \sum_i \frac{w_i}{2}\left\{\tanh\left[ b\left(T_{\rm mid} - T_{\rm gas}\right)\right] + 1  \right\}\,,
\end{equation}
where
\begin{equation}
	b = \frac{\tanh^{-1}\left(2\varepsilon-1\right)}{T_{\rm mid}-T_{\rm max}}\,,
\end{equation}
with $\varepsilon=10^{-3}$, $T_{\rm mid}=(T_{\rm max}+T_{\rm min})/2$, and the sum is over grain species carbon (with parameters $w_i=0.85,\,T_{\rm min}=750\,{\rm K},\,T_{\rm max}=1100\,{\rm K})$, (MgFe)SiO$_4$ $(w_i=0.144,\,T_{\rm min}=1200\,{\rm K},\,T_{\rm max}=1300\,{\rm K})$, and Al$_2$O$_3$ $(w_i=0.006,\,T_{\rm min}=1600\,{\rm K},\,T_{\rm max}=1700\,{\rm K})$.

We tested the validity of our assumptions using \krome{} \citep{Grassi2014}. The initial conditions are given in Tab.\ A2 of M16, and the dust is initially neutral with $n_{\rm d}=1.73\times10^{-10}\,n_{\rm tot}$. In \krome{}, HCO$^{(+)}$ and Mg$^{(+)}$ are used as proxies for m$^{(+)}$ and M$^{(+)}$, and we impose $\dd n_x/\dd t=0$  for the neutrals. We modelled the system until all species reach equilibrium. Empirically, we used $t_{\rm end}=10^6$~yr for models with $n_{\rm tot}<10^{10}$~cm$^{-3}$ and $t_{\rm end}=1$~yr otherwise. We also imposed 
\begin{equation}
	t_{\rm end} =n_{\rm H_2}\times\left(\max\left[\frac{\dd n_{\rm K^+}}{\dd t},\frac{\dd n_{\rm Na^+}}{\dd t},\frac{\dd n_{\rm H^+}}{\dd t}\right]\right)^{-1}\,
\end{equation}
when $n_{\rm tot}\geq10^{20}$~cm$^{-3}$ (i.e.\ when \eqn{eqn:M16_thermoionization} dominates) in order to avoid an over-production of ions given the fact that we impose $\dd n_{\rm H_2} / \dd t=\dd n_{\rm K} / \dd t=\dd n_{\rm Na} / \dd t=0$.

The model results are identical to the results reported in M16.

\begin{table}
\begin{tabular}{llllclll}
\hline
 1 & H$^+$ & + & O & $\to$ & m$^+$ &&\\
 2 & H$^+$ & + & O$_2$ & $\to$ & m$^+$ &&\\
 3 & H$^+$ & + & M & $\to$ & M$^+$ &&\\
 4 & He$^+$ & + & H$_2$ & $\to$ & H$^+$ &&\\
 5 & He$^+$ & + & CO & $\to$ & C$^+$ &&\\
 6 & He$^+$ & + & O$_2$ & $\to$ & m$^+$ &&\\
 7 & H$_3$$^+$ & + & CO & $\to$ & m$^+$ &&\\
 8 & H$_3$$^+$ & + & O & $\to$ & m$^+$ &&\\
 9 & H$_3$$^+$ & + & O$_2$ & $\to$ & m$^+$ &&\\
 10 & H$_3$$^+$ & + & M & $\to$ & M$^+$ &&\\
 11 & C$^+$ & + & H$_2$ & $\to$ & m$^+$ &&\\
 12 & C$^+$ & + & O$_2$ & $\to$ & m$^+$ &&\\
 13 & C$^+$ & + & O$_2$ & $\to$ & m$^+$ &&\\
 14 & C$^+$ & + & M & $\to$ & M$^+$ &&\\
 15 & m$^+$ & + & M & $\to$ & M$^+$ &&\\
 16 & H$^+$ & + & e$^-$ & $\to$ &no products&  & \\
 17 & He$^+$ & + & e$^-$ & $\to$ &no products &&\\
 18 & H$_3$$^+$ & + & e$^-$ & $\to$ &no products & & \\
 19 & H$_3$$^+$ & + & e$^-$ & $\to$ &no products && \\
 20 & C$^+$ & + & e$^-$ & $\to$ &no products  &  & \\
 21 & m$^+$ & + & e$^-$ & $\to$ &no products  &  & \\
 22 & M$^+$ & + & e$^-$ & $\to$ &no products & & \\
\hline
 23 & H$_2$ & & & $\to$ & H$_3$$^+$ & + & e$^-$\\
 24 & He & & & $\to$ & He$^+$ & + & e$^-$\\
 25 & H$_2$ & & & $\to$ & H$^+$ & + & e$^-$\\
\hline
\end{tabular}\caption{List of reactions in UN90 considering the assumptions made (see text). Since the evolution of neutral species is not tracked, they are omitted as products, leading to a lack of mass conservation. Note, however, that the charge is instead correctly conserved. Reactions 1, 6, and 13 assume that O$^+$ is instantaneously converted into OH$^+$, i.e., m$^+$, while reaction 23 assumes that H$_3^+$ is immediately formed from H$_2^+$, and any H produced in reaction 24 is omitted.}
\label{tab:UN90_network}
\end{table}

\section{Initial abundances}\label{sect:initial}
We derive the initial abundances for each species from the total mass density ($\rho$), the dust-to-gas mass ratio ($\mathcal{D}$), and the ionisation fraction ($f_i=n_{\rm e^-} / n_{\rm H_2}$) as follows:
\begin{eqnarray}
 \rho_{\rm H_2} & = & \frac{\rho}{1+\mathcal{D}+f_i\left( m_{\rm e^-}+ m_{\rm Mg^+} \right) / m_{\rm H_2}};\nonumber\\
 \rho_{\rm e^-} & = & f_i \rho_{\rm H_2} \frac{m_{\rm e^-}}{m_{\rm H_2}};\nonumber\\
 \rho_{\rm Mg^+} & = & f_i \rho_{\rm H_2} \frac{m_{\rm Mg^+}}{m_{\rm H_2}};\\
 \rho_{\rm g(Z=0)} & = & \rho_{\rm H_2} \mathcal{D};\nonumber\\
 \rho_{\rm g(Z\ne 0)} & = & 0\nonumber\,,
 \label{eq:chem_init_abunds}
\end{eqnarray}
which guarantees $\rho = \rho_{\rm H_2} + \rho_{\rm e^-} + \rho_{\rm Mg^+} + \rho_{\rm g(Z=0)}$ and global charge conservation.

\section{Cooling time}\label{sect:cooling_time}
To compute the cooling time of a static volume of gas, we first derive its temperature evolution given an initial temperature $T(t=0)$, assuming a density $n_{\rm tot}\simeq n_{\rm H_2}$ and a cosmic ray ionisation rate $\zeta$.
By equating \eqn{eqn:pressure} and \eqn{eqn:pressure_temperature}, we find
\begin{equation}
 T = (\gamma-1)\frac{E-E^*}{n_{\rm tot}k_{\rm B}}\,,
\end{equation}
where $E^*$ is the sum of kinetic and magnetic energy, and we assume both to be constant in time for the current purpose. The time derivative of the temperature is thus
\begin{equation}
 \frac{\dd T}{\dd t} = \frac{\dd E}{\dd t} \frac{\gamma-1}{n_{\rm tot}k_{\rm B}}\,.
\end{equation}
Since we are considering a static volume of gas, i.e., all quantities are spatially constant, from \eqn{eqn:hydro_full_energy} we obtain $\partial_t E=-\Lambda_{\rm chem} + \Gamma_{\rm CR}$, followed by
\begin{equation}
 \frac{\dd T}{\dd t}=(\gamma-1) \frac{\Gamma_{\rm CR}-\Lambda_{\rm chem}(T)}{n_{\rm tot} k_{\rm B}}\,.
\end{equation}

\bsp

\label{lastpage}

\end{document}